\documentclass[aps,prd,reprint,superscriptaddress,nofootinbib,longbibliography,floatfix,raggedbottom]{revtex4-2}
\usepackage[T1]{fontenc}
\usepackage[utf8]{inputenc}
\usepackage{lmodern}
\usepackage{amsmath,amssymb,bm,booktabs,array,graphicx,xcolor}
\usepackage{tikz}
\usepackage{placeins}
\usetikzlibrary{arrows.meta,positioning,fit,shapes.misc}

\usepackage[hidelinks,unicode]{hyperref}
\hypersetup{pdftitle={ASTRA: Action-based stochastic transport for lattice field theory},pdfsubject={Boltzmann sampling and lattice field theory}}
\graphicspath{{figures/}}
\newcommand{\dd}{\mathrm d}
\newcommand{\E}{\mathbb E}
\newcommand{\Var}{\operatorname{Var}}
\newcommand{\Cov}{\operatorname{Cov}}

\newcommand{\ESS}{\operatorname{ESS}}

\newcommand{\R}{\mathbb R}
\newcommand{\Z}{\mathbb Z}

\makeatletter
\newcounter{algorithm}\providecommand{\theHalgorithm}{\arabic{algorithm}}
\def\fps@algorithm{tbp}\def\ftype@algorithm{4}\def\ext@algorithm{loa}
\def\fnum@algorithm{\algorithmname~\thealgorithm}\newcommand{\algorithmname}{Algorithm}
\newenvironment{algorithm}[1][tbp]{\@float{algorithm}[#1]\hrule height 0.8pt\kern4pt}{\par\kern2pt\hrule height 0.8pt\end@float}
\makeatother
\usepackage{algpseudocode}

\begin{document}
\title{ASTRA: Action-based stochastic transport for lattice field theory}
\author{Moxian Qian}
\email{mqian@students.uni-mainz.de}
\affiliation{Johannes Gutenberg University Mainz, 55128 Mainz, Germany}
\author{Shiyang Chen}
\email{2222895@swansea.ac.uk}
\affiliation{Centre for Quantum Fields and Gravity, Department of Physics, Swansea University, Swansea SA2 8PP, United Kingdom}
\author{Yuanqi Du}
\email{yuanqidu@microsoft.com}
\affiliation{Microsoft Research New England, Cambridge, Massachusetts 02142, USA}

\begin{abstract}
Lattice simulations predict the properties of quantum fields by averaging over their possible configurations. In lattice gauge theories on fine lattices, a simulation can remain trapped in one topological sector, obscuring the charge fluctuations it aims to measure. In this paper, we introduce action-based stochastic transport (ASTRA), which directly transports configurations from a simple initial distribution to the target, bypassing the long mixing time of mode exploration in conventional Markov chain Monte Carlo algorithms. ASTRA instead iteratively trains on its own simulated configurations, corrected by sequential Monte Carlo with path importance weights, with no equilibrium training data needed. Furthermore, the design of ASTRA accommodates transferability across lattice sizes and coupling strengths. In the two-dimensional compact $U(1)$ gauge theory, ASTRA trained at $L=8$ with couplings from $\beta=6$ to $96$ samples lattices up to $L=32$ and $\beta=96$ without retraining, including couplings at which reference Hamiltonian Monte Carlo run directly at the target coupling never changes the topological charge. ASTRA reproduces the exact topological susceptibility, its continuum limit, and Wilson loops.
The same construction carries over to non-Abelian gauge fields, with the noise acting in the Lie algebra of the gauge group. In two-dimensional $SU(3)$ gauge theory, a model trained at $L=8$ for couplings from $\beta=6$ to $16$ samples lattices up to $L=32$ without retraining. The model reproduces the exact Wilson loops up to area $16$.
\end{abstract}

\maketitle

\section{Introduction}
\label{sec:intro}

Lattice field theory expresses nonperturbative observables as statistical averages by discretizing Euclidean spacetime~\cite{Wilson1974}.
Reliable continuum predictions require simulations at several lattice spacings, control of statistical and systematic errors, and an extrapolation to the continuum limit. As the lattice spacing decreases at fixed physical correlation length, the correlation length in lattice units grows. Conventional algorithms then suffer from critical slowing down, and their autocorrelation times increase. Effectively independent configurations become more expensive. This cost limits the accessible lattice spacings and volumes and the attainable precision~\cite{Schaefer2011}. More efficient sampling algorithms are therefore needed.

Lattice gauge theories add the problem of topological freezing. Markov chain Monte Carlo (MCMC) changes a configuration by small local updates. Near the continuum limit these updates almost never change the topological charge, particularly with periodic boundary conditions~\cite{DelDebbio2004,Luscher2011}. Chains then stay in one topological sector for a long time. The topological susceptibility, which measures the fluctuations of the charge, is affected most, and other observables that depend on the charge are biased as well~\cite{Aoki2007FixedTopology}.

Several Monte Carlo methods mitigate these problems. Against critical slowing down, cluster algorithms update correlated degrees of freedom collectively, and Fourier acceleration reduces the differences in relaxation time between momentum modes~\cite{Brower1989,Christ2018,Ostmeyer2025}. Other approaches target topological freezing. Open boundary conditions allow topological charge to flow through the boundaries~\cite{Luscher2011}, whereas parallel tempering in boundary conditions employs auxiliary replicas to facilitate transitions between topological sectors while preserving a periodic target ensemble~\cite{Hasenbusch2017}. Further methods include winding updates, which propose topology-changing transitions with an acceptance correction~\cite{Albandea2021}, and metadynamics, which introduces bias potentials along selected collective variables~\cite{Laio2016,Eichhorn2024}. Parallel tempering in boundary conditions has also been extended to full quantum chromodynamics (QCD) at physical quark masses~\cite{Bonanno2024}.

Generative models based on deep learning have more recently been applied to both problems. For critical slowing down, normalizing flows (NFs), stochastic normalizing flows (SNFs), and continuous normalizing flows (CNFs) learn transformations or stochastic dynamics that give more efficient sampling distributions~\cite{Albergo2019,Caselle2022,Bacchio2023,Bulgarelli2025,Singha2026}. Diffusion models and the Stochastic Path Sampler (SPS) learn stochastic sampling paths and have been applied to scalar field theories near criticality~\cite{Wang2024SQ,Chen2026SPS}. In lattice gauge theories, gauge-equivariant NFs and SNFs explore topological sectors better, and CNFs have been studied as flexible architectures for gauge fields~\cite{Kanwar2020,Boyda2021,Gerdes2024,Bonanno2026Boundary}. Diffusion-based approaches have also shown promising results in mitigating topological freezing in two-dimensional compact $U(1)$ gauge theory~\cite{Zhu2026Conditioned,Vega2026}. Diffusion models for non-Abelian gauge theories have been trained on Monte Carlo ensembles and applied to other couplings and larger lattices~\mbox{\cite{Aarts2026Generalizable,Komijani2026SUN}}, and data-free training of diffusion models has been proposed for lattice field theories~\mbox{\cite{Grinwald2026,Tomiya2026}}. 

Since the learned distribution generally mismatches the target distribution, generative samplers are typically combined with Metropolis-Hastings (MH) chains, which correct for this mismatch and allow autocorrelations to be assessed. Normalizing flows serve as proposals in independence Metropolis-Hastings (IMH) chains~\cite{Albergo2019,Kanwar2020}, while variational autoregressive networks regenerate ordered blocks of lattice sites within an MH update~\cite{Qian2025Block}. For stochastic path samplers, whose endpoint densities are intractable, neural nonequilibrium Hamiltonian Monte Carlo (NHMC) introduces round-trip (RT) MH chains along nonequilibrium paths~\cite{Qian2026NHMC}. SPS applies one IMH step to the whole path~\cite{Chen2026SPS}, and NHMC applies RT updates to it. Both train the paths with a variational objective in path space. The acceptance of a whole path falls as the variance of the log weight grows with the volume~\cite{Qian2026NHMC}.
Learned models can also be embedded in other exact algorithms. Neural transports have been used to accelerate parallel tempering~\cite{Zhang2025APT}, and in lattice field theory a time-dependent energy learned by score matching drives flow-transformed replica exchange (T-REX), where the same trained model can also be used in stochastic normalizing flows~\cite{KanwarVega2026}.  An energy learned by score matching has also been used for thermodynamic integration for free energy estimation in Lennard-Jones fluids~\cite{Mate2024NTI}. 

These generative models are trained in two ways. Data-free variational training uses the reverse Kullback--Leibler (KL) divergence, defined either in configuration space or in path space~\cite{Albergo2019,Wu2019VAN,DelDebbio2021,Chen2026SPS}. Its gradients come from samples of the model itself, so a mode that the model rarely visits receives little training signal and can be lost. This mode collapse is a known failure of reverse-KL training~\cite{Nicoli2023,Hackett2025,Soletskyi2025}. Data-driven training uses the forward KL divergence~\cite{Noe2019,Hackett2025,Wang2024SQ}. It covers the modes present in the data, but it needs representative equilibrium configurations, which are exactly what critical slowing down and topological freezing make expensive. If the data cover the modes poorly, the learned relative weights of the modes are also wrong~\cite{Hackett2025,Wenliang2021}. A sampler that corrects its own samples avoids both problems. It needs no precomputed equilibrium data, and the correction, rather than the model, sets the relative weights of the modes that the samples reach.

We introduce action-based stochastic transport (ASTRA), which rests on three ideas. First, ASTRA transports a population of configurations (i.e., particles) from a simple initial distribution to the target in a finite number of steps. The drift of the transport combines a physics term computed from the action with a learned residual. Changes of the topological charge happen early along the transport, where the added noise weakens the coupling and local updates mix quickly. Second, ASTRA learns while it simulates. The transport can be corrected via sequential Monte Carlo (SMC)~\cite{Neal2001,DelMoral2006}, where path importance weights~\cite{Jarzynski1997,Crooks1999} correct the particles between intermediate times, resampling copies them in proportion to these weights, and Hamiltonian Monte Carlo (HMC) moves mix them at each time. The corrected configurations then train the model of the next round, with the force from the action as a regression target~\cite{DeBortoli2024}. 

Third, the learned residual is local and conditioned on the coupling. One model trained on a small lattice is therefore applied to larger lattices and to other couplings without retraining.

In two-dimensional compact $U(1)$ gauge theory, one model trained at $L=8$ is applied unchanged to lattices up to $L=32$ and couplings up to $\beta=96$. At these couplings, reference HMC run directly at the target coupling never changes the topological charge in $4\times10^5$ trajectories on the larger lattices. Along two lines of constant physics (fixed $\beta/L^2$), ASTRA reproduces the exact topological susceptibility at every lattice size, and its continuum extrapolation agrees with the exact value. Wilson loops at $L=32$ also agree with the exact results, whereas reference HMC with a hot start, that is, started from random links,
underestimates them. Two-dimensional $SU(3)$ gauge theory tests ASTRA on a non-Abelian group, where each link is a $3\times3$ unitary matrix and the noise acts in the eight-dimensional Lie algebra. A model trained at $L=8$ reproduces the exact Wilson loops up to $L=32$ without retraining, so the transfer carries over from the Abelian to the non-Abelian case. An asymmetric double well isolates the question of mode coverage. There all five trained ASTRA models retain both modes, whereas three of five models trained by reverse KL in path space lose one.

The remainder of this paper is organized as follows. Section~\ref{sec:objectives} presents ASTRA, its training, and its transfer across lattice sizes. Section~\ref{sec:numerical} tests mode coverage in the double well. Section~\ref{sec:u1} studies topological sampling in two-dimensional $U(1)$ gauge theory. Section~\ref{sec:su3} applies ASTRA to the non-Abelian $SU(3)$ gauge theory in two dimensions, where a model trained at $L=8$ is used up to $L=32$, and Sec.~\ref{sec:conclusion} concludes.

\begin{figure*}[t]
\centering\includegraphics[width=\textwidth]{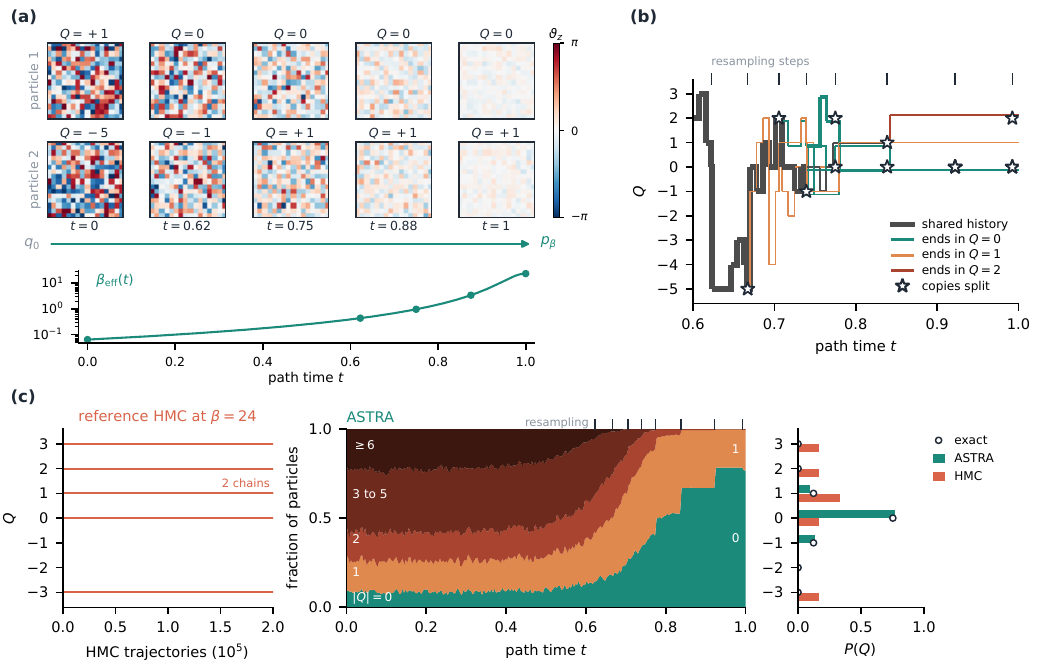}
\caption{One ASTRA sampling run for two-dimensional $U(1)$ gauge theory at $L=16$ and $\beta=24$, with a model trained at $L=8$ ($N=1024$ particles, that is, lattice configurations, and $T=4096$ path steps). The path time $t$ runs from random links, drawn from $q_0$, at $t=0$ to the target $p_\beta$ at $t=1$. The particles change their topological charge $Q$ freely while the coupling is weak, $Q$ settles by $t\approx0.8$, and the run ends close to the exact distribution of $Q$, whereas reference HMC at $\beta=24$ stays in its initial sectors. (a) Plaquette angles $\vartheta_z$ along the ancestral lines of two final particles, one ending in $Q=0$ and one in $Q=1$, at five path times, with $Q$ above each map. The maps at $t=0.62$ follow the first resampling step. Below, the coupling $\beta_{\rm eff}(t)$ of the physics term [Eq.~\eqref{eq:u1_beta_eff}] rises from $0.06$ to $\beta$ and sets the annealing schedule along the path. (b) $Q$ for $t\geq0.6$ along the histories of the $26$ final descendants of the initial particle whose descendants end in the most sectors. Before the first resampling step at $t=0.62$ they share one history. A history shared by descendants that end in different sectors is dark gray, with a width proportional to the number of descendants that share it, and a history whose descendants all end in one sector has the color of that sector (teal, orange, and brick for $Q=0$, $1$, and $2$), slightly offset so that histories ending in different sectors do not overlap. Stars mark resampling steps (ticks) at which copies split. Between resampling steps $Q$ changes through the propagation steps, by several units at once while the coupling is weak, and at resampling steps also through the HMC moves. (c) Left, six reference HMC chains at $\beta=24$ started from random links stay in their initial sectors over $2\times10^5$ trajectories, and two of them share $Q=1$. Middle, unweighted fraction of ASTRA particles with each $|Q|$ along the path, with ticks at resampling steps. Right, the weighted $P(Q)$ of this run and the fraction of HMC chains in each sector, compared with the exact values.}
\label{fig:framework}
\end{figure*}
\section{Action-based stochastic transport}
\label{sec:objectives}

ASTRA alternates between simulation and learning. In each training round the current model transports particles to the target, the SMC correction turns them into configurations that follow the target, and these configurations train the model by force regression. After training the model is frozen. Sampling runs, the runs used for measurements, apply the same SMC transport to independent populations. Figure~\ref{fig:framework} follows one sampling run in two-dimensional $U(1)$ gauge theory, from random links at $t=0$ to the target at $t=1$. Section~\ref{sec:astra_production} describes the transport and its SMC correction, Sec.~\ref{sec:astra_training} the training, and Sec.~\ref{sec:astra_transfer} the local model that transfers across lattice sizes and couplings.

Let $x$ denote a complete field configuration with action $S_\beta(x)$. The target Boltzmann distribution is
\begin{equation}
p_\beta(x)=\frac{e^{-S_\beta(x)}}{Z_\beta},
\qquad
Z_\beta=\int e^{-S_\beta(x)}\,\dd\mu(x),
\label{eq:astra_target}
\end{equation}
where $\dd\mu$ denotes the reference measure, given by the Lebesgue measure for Euclidean variables or the product of normalized Haar measures for gauge links. For a lattice field theory, $S_\beta$ is the Euclidean action, the discretized integral of the Lagrangian density, and $e^{-S_\beta}$ is the Boltzmann weight of a configuration.

\subsection{Transport and SMC correction}
\label{sec:astra_production}

Instead of learning the whole drift, ASTRA adds a learned residual to the physical action. The time-dependent effective action is 
\begin{equation}
S_\theta(x,t;\beta)
=
S_{\mathrm{phys}}(x,t;\beta)-\mathcal E_\theta(x,t;\beta),
\label{eq:astra_effective_action}
\end{equation}
where $S_{\mathrm{phys}}$ is constructed from the known physical action and $\mathcal E_\theta$ is the learned residual. The resulting drift is
\begin{equation}
s_\theta(x,t;\beta)
=
-\nabla_x S_\theta(x,t;\beta).
\label{eq:astra_proposal_force}
\end{equation}
The physics term is fixed analytically, and the residual learns corrections to the transport. ASTRA learns a scalar effective action rather than the drift itself. The path weights and the HMC moves at intermediate times below need the intermediate density $e^{-S_\theta}$, and its gradient then gives the drift. A scalar also makes the drift a gradient field.

Paths start from a tractable reference distribution $q_0$ and follow a discretized diffusion driven by the drift. Here $t\in[0,1]$ is the path time, $\sigma(t)$ is the noise scale, and $x$ has $d$ real components. With $T$ path steps, $\Delta t=1/T$, $t_k=k/T$, and $v_k=\Delta t\,g^2((t_k+t_{k+1})/2)$, the forward transition is
\begin{equation}
x_{k+1}
=
x_k+v_k s_\theta(x_k,t_k;\beta)
+\sqrt{v_k}\,\xi_k,
\qquad
\xi_k\sim\mathcal N(0,I_d),
\label{eq:astra_forward}
\end{equation}
where $g^2(t)=-\dd\sigma^2(t)/\dd t$. The transition defines a tractable forward kernel $f_k(x_{k+1}\mid x_k)$, and we call one application of Eq.~\eqref{eq:astra_forward} a propagation step.

We introduce positive unnormalized intermediate densities
\begin{equation}
\gamma_k(x)=e^{-S_\theta(x,t_k;\beta)}.
\label{eq:astra_intermediate}
\end{equation}
For the learned intermediate density, the drift satisfies $s_\theta(x,t_k;\beta)=\nabla_x\log\gamma_k(x)$. The corresponding HMC moves preserve the normalized distribution associated with $\gamma_k$. Other intermediate densities can be used if the update kernels preserve them.

Let $f_k(x_{k+1}\mid x_k)$ denote the forward transition density and $b_k(x_k\mid x_{k+1})$ an auxiliary normalized backward transition density. The initial importance weight is
\begin{equation}
w_0(x_0)=\frac{\gamma_0(x_0)}{q_0(x_0)},
\label{eq:astra_w0}
\end{equation}
and the incremental path weight is
\begin{equation}
G_k=
\frac{
\gamma_{k+1}(x_{k+1})
b_k(x_k\mid x_{k+1})
}{
\gamma_k(x_k)
f_k(x_{k+1}\mid x_k)
}.
\label{eq:astra_increment}
\end{equation}
The marginal density of a transported particle at time $t_k$ is an integral over all earlier paths and cannot be efficiently evaluated. The weights are therefore defined on whole paths, and the backward kernel $b_k$ is an auxiliary reverse transition that makes this path ratio computable~\cite{DelMoral2006}. The backward kernel is a choice of the method. Any normalized $b_k$ gives correct weights, and a $b_k$ closer to the time reversal of the forward dynamics gives a smaller weight variance. We use a Gaussian with the same variance as the forward kernel and zero drift. It coincides with the time reversal of a forward step without drift, and its Gaussian transition density $b_k(x_k\mid x_{k+1})$ has a closed form.

At each intermediate stage, the path weights are updated using $G_k$. Their concentration is monitored through the importance-weight effective sample size (ESS)
\begin{equation}
\ESS_w
=
\frac{1}{\sum_{i=1}^{N}(\bar w_i)^2},
\qquad
\bar w_i=\frac{w_i}{\sum_j w_j}.
\label{eq:astra_ess}
\end{equation}
When $\ESS_w/N<1/2$, the population is resampled systematically according to the weights. Short HMC trajectories targeting the current intermediate density, which we call HMC moves, then separate the duplicated particles and improve exploration. The HMC accept-reject step preserves the corresponding intermediate density. Weights and resampling thus move the population between times, and HMC moves mix it within one time. Together they form a resample-move SMC sampler~\cite{DelMoral2006}, summarized in Algorithm~\ref{alg:production}.

Upon reaching the terminal time, the particles receive a final correction
\begin{equation}
w_T^{\mathrm{final}}
\longleftarrow
w_T\,
\frac{e^{-S_\beta(x_T)}}{\gamma_T(x_T)}.
\label{eq:astra_terminal}
\end{equation}
It corrects for the difference between the last intermediate density and the physical target. Without resampling and intermediate Markov updates, the full path weight is
\begin{equation}
w_{\mathrm{path}}(x_{0:T})
=
\frac{
e^{-S_\beta(x_T)}
\prod_{k=0}^{T-1}b_k(x_k\mid x_{k+1})
}{
q_0(x_0)
\prod_{k=0}^{T-1}f_k(x_{k+1}\mid x_k)
}.
\label{eq:astra_path_weight}
\end{equation}
The weight uses the discretized kernels of the actual simulation.

Finally, an observable $O$ is estimated from the terminal weighted particles as
\begin{equation}
\widehat O
=
\sum_{i=1}^{N}
\bar w_i^{\mathrm{final}}O(x_T^{(i)}),
\qquad
\bar w_i^{\mathrm{final}}
=
\frac{w_i^{\mathrm{final}}}
{\sum_j w_j^{\mathrm{final}}}.
\label{eq:astra_observable}
\end{equation}
Independent sampling runs give the statistical errors and test for effects of the finite population. Path weights correct toward the target, but a finite population and resampling can still add bias and variance.

A sampling run applies this procedure with the frozen model to its own population of $N$ particles drawn from $q_0$ [Fig.~\ref{fig:framework}(a)], and the network is not updated (Algorithm~\ref{alg:production}).

\begin{algorithm}[!htbp]
\caption{One sampling run with a frozen model.}
\label{alg:production}
\begin{algorithmic}[1]
\Require frozen model; $N$ particles; $T$ path steps; ESS threshold
\State draw independent $x_i\sim q_0$, set $\ell_i=\log\gamma_0(x_i)-\log q_0(x_i)$, and assign distinct labels of initial ancestors
\For{$k=0,\dots,T-1$}
  \State propagate with Eq.~\eqref{eq:astra_forward}; add $\log G_k$, computed from the coordinates actually generated, before angles are wrapped (Appendix~\ref{app:compact}), to $\ell_i$
  \State normalize the weights
  \If{the ESS threshold is reached}
    \State systematic resampling; copy ancestor labels; reset the relative log weights to zero
  \EndIf
  \If{resampling occurred at step $k$}
    \State apply HMC moves invariant under $\gamma_{k+1}$
  \EndIf
\EndFor
\State add $-S_\beta(x_T)-\log\gamma_T(x_T)$ to $\ell_i$
\State estimate observables with the normalized final weights; save the estimate and the runtime
\end{algorithmic}
Independent runs are combined with Eq.~\eqref{eq:run-stats}.
\end{algorithm}

\subsection{Training}
\label{sec:astra_training}

In each training round the current model generates independent paths from $q_0$ with Eq.~\eqref{eq:astra_forward}, and their path weights, Eq.~\eqref{eq:astra_path_weight}, correct them at the endpoint. The endpoints are resampled with these weights and updated by HMC moves targeting $p_\beta$. In early rounds the weights are very uneven, so resampling keeps only a few distinct paths, and the HMC moves spread their copies. Symmetry transformations of the target are also applied. These steps give the corrected configurations for training.

A replay buffer stores the corrected configurations of all training rounds. For a configuration $X$ from the buffer, the noise kernel gives a noisy input $Y_t$. The physical force
\begin{equation}
s_\beta(X)=-\nabla_X S_\beta(X)
\label{eq:astra_physical_force}
\end{equation}
is evaluated at $X$ and is the regression target for the drift at $Y_t$. The training objective is
\begin{equation}
\mathcal L_{\mathrm{force}}(\theta)
=
\mathbb E_{X\sim r,\;Y_t\sim \mathcal K_t(\cdot\mid X),\;t}
\left[
\left\|
s_\theta(Y_t,t;\beta)-s_\beta(X)
\right\|^2
\right],
\label{eq:astra_force_loss}
\end{equation}
where $r$ is the distribution of the replay buffer and $\mathcal K_t$ is the noise kernel. This is a fixed-point iteration based on target score matching~\cite{DeBortoli2024,Phillips2024,pmlr-v306-blessing26a}. 
If the training configurations follow the target and regularity conditions hold, the optimal regression function is the score of the smoothed target. In practice the regression target depends on the buffer distribution, so the quality and mode coverage of the corrected configurations matter. The resulting shift of the learned drift is given in Sec.~S1 of the Supplemental Material. It is small when the noisy modes remain separated and grows where they overlap.

Training thus alternates between transport, correction, and force regression. The model improves on configurations produced by the sampler itself, without a precomputed equilibrium dataset.

\begin{algorithm}[!htbp]
\caption{One ASTRA training round.}
\label{alg:training}
\begin{algorithmic}[1]
\Require current model; training couplings $\{\beta\}$; replay buffer
\For{each training coupling $\beta$}
  \State generate independent paths from $q_0$ with Eq.~\eqref{eq:astra_forward} and compute their path weights
  \State resample the endpoints with the path weights and apply HMC moves targeting $p_\beta$
  \State add the corrected configurations and their symmetry images to the buffer
\EndFor
\State fit $s_\theta(Y_t)$ to $s_\beta(X)$ by minimizing Eq.~\eqref{eq:astra_force_loss} over the buffer
\end{algorithmic}
\end{algorithm}
 Appendix~\ref{app:dw_reweight} and Sec.~S3 of the Supplemental Material compare endpoint resampling with other ways of using the path weights in training.

Training on corrected configurations and then sampling with the improved model is closely related to the particle denoising diffusion sampler (PDDS)~\mbox{\cite{Phillips2024}} and to other samplers that combine learned transport with SMC~\mbox{\cite{Matthews2022,Midgley2023,Cheng2026}}. ASTRA adds a physics term to the learned residual and uses a local model that transfers across lattice sizes.

\subsection{Local model and transfer}
\label{sec:astra_transfer}

The same frozen model serves every lattice size and coupling. The learned residual is a sum over sites of an energy density. A network of local convolutional layers on the periodic lattice, with a finite receptive field, computes this density from gauge-invariant local features, and the time and the coupling enter through gates that scale the features of each layer. The parameters therefore do not depend on $L$. A model trained at $L=8$ is evaluated at $L=32$ by applying the same network to the larger lattice, and couplings between the training values are reached through the conditioning on $\beta$. Only the HMC step size and the number of resampling steps, which the resampling rule of Sec.~\ref{sec:astra_production} sets adaptively, change with $L$.

\section{Asymmetric double well}
\label{sec:numerical}

Before turning to gauge fields, we test on an asymmetric double well whether fitting the force to corrected configurations keeps the modes of a multimodal target, which reverse-KL training can lose (Sec.~\ref{sec:intro}).

The action is
\begin{equation}
S_{\mathrm{DW}}(x)
=
a(x^2-b^2)^2+hx,
\qquad a>0,\quad b>0,
\label{eq:double_well_action}
\end{equation}
where $a$ controls the barrier height, $b$ determines the locations of the two wells, and $h$ introduces an asymmetry between their probabilities. A second coordinate $y$ with the independent Gaussian action $y^2/2$ is added, so the sampler works in two dimensions. Let $P_R$ be the probability that $x>0$. We use $a=0.5$, $b=2.5$, and $h=-0.3$, for which numerical quadrature gives $P_R=0.8153$. A single sampling run gives $\widehat P_R=\sum_i\bar w_i^{\mathrm{final}}\mathbf1_{\{x_T^{(i)}>0\}}$, and $\overline P_R$ denotes its mean over the sampling runs of one trained model.

The double well has no physics term. The part of the intermediate density that does not depend on the model is a Gaussian, $\exp[-|x|^2/(2\sigma^2(t)+2)]$, which carries no information about the target, and the learned residual enters as the factor $e^{\mathcal E_\theta}$ that multiplies it.

\begin{figure}[!tp]
\centering\includegraphics[width=\columnwidth]{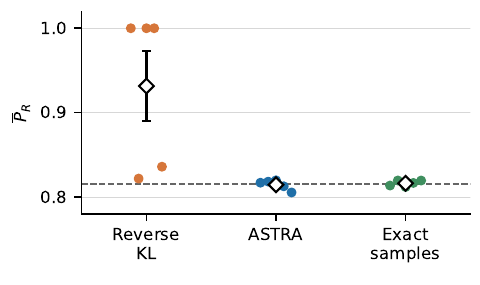}
\caption{Mode coverage in the double well. Each circle is one trained model, with $\overline P_R$ averaged over its $16$ sampling runs, and diamonds give the mean of the five models with its standard error. The models are trained from scratch by reverse KL in path space or by ASTRA, whose training here also resamples at intermediate times (Appendix~\ref{app:doublewell_results}), and as a reference fitted with the ASTRA loss to exact target samples (Exact samples). A model with $\overline P_R=1$ has lost the left well. The dashed line is the exact $P_R=0.8153$. }
\label{fig:double_well}
\end{figure}

Figure~\ref{fig:double_well} compares five models trained from scratch by ASTRA with five trained by reverse KL in path space. The baseline uses the same family of paths, and its reparameterized gradients also use the physical force, so the two differ only in the objective and in the configurations on which the loss is evaluated. At the start both wells are populated. Three of the five reverse-KL models then collapse to the right well and give $\overline P_R=1$. All five ASTRA models keep both wells in every sampling run, with estimates around $P_R=0.8153$, as do models fitted with the same loss to exact samples. The loss of Eq.~\eqref{eq:astra_force_loss} fits the force on the corrected configurations, and the correction keeps the relative weight of each well among them. A well therefore stays in the fit as long as the corrected configurations reach it.

The same mechanism cannot bring back a well that the corrected configurations no longer reach. A model that has lost the left well has almost no density there, and the Gaussian factor cannot restore it, because it multiplies $e^{\mathcal E_\theta}$. Five models first fitted to configurations of the right well only therefore stay collapsed under Algorithm~\ref{alg:training}. If instead $10\%$ of the paths follow the drift of the Gaussian alone, the path density becomes a mixture, which has weight in both wells. With every path weighted by this mixture, all five models regain the left well and give $\overline P_R=0.813$ (Appendix~\ref{app:dw_recovery}). ASTRA thus keeps the modes that its corrected configurations reach, but it does not undo a collapse by itself.

Reusing corrected configurations in a replay buffer further stabilizes the fit (Appendix~\ref{app:dw_replay}), and Appendix~\ref{app:dw_reweight} compares other ways of using the path weights. Section~S5 of the Supplemental Material applies ASTRA to two-dimensional $\phi^4$ theory, where the two branches of the magnetization play the role of the wells. For gauge fields the corresponding question is whether the corrected configurations reach all topological sectors, which Sec.~\ref{sec:u1} examines.

\section{\texorpdfstring{ASTRA for two-dimensional $U(1)$ gauge theory}{ASTRA for two-dimensional U(1) gauge theory}}
\label{sec:u1}

We consider two-dimensional compact $U(1)$ lattice gauge theory on an $L\times L$ square lattice with periodic boundary conditions. The lattice contains $V=L^2$ sites and plaquettes, with $d=2V$ oriented links. The gauge variables are elements of $U(1)$, parameterized as $U_{z,\mu}=e^{ix_{z,\mu}}$, where $x_{z,\mu}\in[-\pi,\pi)$, $z$ labels lattice sites, and $\mu=0,1$ denotes the two positive lattice directions. The oriented plaquette angle is defined by

\begin{equation}
\vartheta_z(x)
=
x_{z,0}+x_{z+\hat{0},1}
-x_{z+\hat{1},0}-x_{z,1}.
\label{eq:plaquette}
\end{equation}
The Wilson gauge action is given by
\begin{equation}
S_\beta(x)
=
-\beta\sum_z\cos\vartheta_z(x),
\label{eq:u1_action}
\end{equation}
where $\beta$ denotes the inverse gauge coupling. This action differs from the conventional expression $\beta\sum_z(1-\cos\vartheta_z)$ only by an additive constant and therefore defines the same normalized Boltzmann distribution.

The paths follow the effective action of Eq.~\eqref{eq:astra_effective_action}, which combines the physics term with the learned residual,
\begin{equation}
S_\theta(x,t;\beta)
=
-\beta_{\mathrm{eff}}(t)
\sum_z\cos\vartheta_z(x)
-\mathcal E_\theta(\vartheta(x),t;\beta),
\label{eq:u1_effective_action}
\end{equation}
where
\begin{equation}
\beta_{\mathrm{eff}}(t)
=
\frac{\beta}{1+4\beta\sigma^2(t)},
\qquad
\sigma(t)=\sigma_{\max}(1-t).
\label{eq:u1_beta_eff}
\end{equation}
Here $t\in[0,1]$ denotes the path time, with the physical target at $t=1$, and $\sigma_{\max}=2$ in our $U(1)$ simulations. The first term is the physics term. It is the Wilson action with the coupling reduced to $\beta_{\mathrm{eff}}(t)$ by the noise, and it is computed from the action without training. Appendix~\ref{app:weakfield} derives $\beta_{\mathrm{eff}}(t)$ from the plaquette noise, and Sec.~S2 of the Supplemental Material gives the momentum dependence it misses. The residual $\mathcal E_\theta$ is a sum of local network outputs built from periodic plaquette features, which makes it gauge invariant. Charge conjugation is imposed by symmetrizing the learned residual. Untrained ASTRA sets $\mathcal E_\theta=0$ and keeps the rest of the construction, so it is ASTRA without learning, and Appendix~\ref{app:no-analytic} compares training with and without the physics term. All $U(1)$ results of the main text use a single model conditioned on $\beta$, trained only at $L=8$ at the six couplings $\beta\in\{6,13.5,24,37.5,54,96\}$ (Appendix~\ref{app:production}). The second line of constant physics in Sec.~\ref{sec:u1_accuracy} and the Wilson loops at $\beta=7$, $8$, $10$, and $12$ use couplings between these training values. The density $\gamma_\theta=e^{-S_\theta}$ is unnormalized, and $S_\theta$ also serves as the HMC potential. The resulting drift is
\begin{equation}
s_\theta(x,t;\beta)
=
-\nabla_x S_\theta(x,t;\beta),
\label{eq:u1_proposal_force}
\end{equation}
which drives the paths. With $\vartheta=Cx$, where $C$ is the plaquette-link incidence matrix, its expression in link coordinates is given in Appendix~\ref{app:compact}.

We use the geometric topological charge
\begin{equation}
Q(x)
=
\frac{1}{2\pi}
\sum_z
\operatorname{Arg}\!\left(e^{i\vartheta_z(x)}\right)
\in\mathbb Z.
\label{eq:topology}
\end{equation}
Periodic boundary conditions ensure that $Q$ is integer-valued, apart from configurations on branch boundaries of measure zero. The probability of a topological sector $q$ is
\begin{equation}
P(Q=q)
=
\int
\mathbf{1}_{\{Q(x)=q\}}
p_\beta(x)\,d\mu(x).
\label{eq:sector_probability}
\end{equation}
The topological susceptibility is defined as
\begin{equation}
\chi_{\mathrm{top}}
=
\frac{
\langle Q^2\rangle_{p_\beta}
-\langle Q\rangle_{p_\beta}^2
}{V}.
\label{eq:top-susceptibility}
\end{equation}
Charge-conjugation symmetry implies $\langle Q\rangle_{p_\beta}=0$, so we evaluate $\chi_{\mathrm{top}}=\langle Q^2\rangle_{p_\beta}/V$.

We also measure Wilson loops to assess the accuracy of gauge-invariant observables beyond topology. For a rectangular region $\mathcal R_{r,s}(z)$ containing $A=rs$ plaquettes, the translation-averaged real Wilson loop is
\begin{equation}
W_{r\times s}(x)
=
\frac{1}{V}\sum_z
\cos\!\left(
\sum_{y\in\mathcal R_{r,s}(z)}
\vartheta_y(x)
\right).
\label{eq:loop-observable}
\end{equation}
We use $W_A$ when specifying only the enclosed area, with the mean plaquette given by $W_{1\times1}$. Exact finite-volume reference formulas for these observables are provided in Appendix~\ref{app:reference}.

\subsection{Sampling and evaluation}
\label{sec:cost}

A sampling run moves $N$ particles with Eq.~\eqref{eq:astra_forward} on the link angles, with the $\sigma(t)$ above. The drift, the weights, and the HMC moves all use the learned intermediate density $\gamma_k$ of Eq.~\eqref{eq:astra_intermediate}, and untrained ASTRA uses the same construction with $\mathcal E_\theta=0$. When $\ESS_w/N<1/2$, the population is resampled systematically~\mbox{\cite{Douc2005,Gerber2020}} and then updated by HMC moves~\mbox{\cite{Duane1987}} with potential $-\log\gamma_{k+1}$, which makes ASTRA a resample-move SMC sampler~\mbox{\cite{DelMoral2006}}. Resampling only copies existing particles, whereas propagation and HMC moves create new configurations and can enter new sectors. Appendix~\ref{app:compact} shows that the path weights correct the population toward $p_\beta$ on compact link angles, and Appendix~\ref{app:production} lists the step sizes and the number of moves.

Particles of one run share ancestors and are not independent. Statistical errors therefore come from independent sampling runs, each with its own population. For $R$ runs with estimates $\widehat O_j$, the mean $\overline O$ has the standard error $s_O/\sqrt R$, where $s_O$ is the sample standard deviation of the $\widehat O_j$ (Appendix~\ref{app:run-stats}).

When the target variance is known (Appendix~\ref{app:reference}), we compare a run with independent target samples through the observable effective sample size
\begin{equation}
 N_{\mathrm{eff}}[O]=\frac{\Var_{p_\beta}(O)}{\Var(\widehat O)},
 \label{eq:obs-efficiency}
\end{equation}
 with the denominator estimated by $s_O^2$. Independent target samples give $N_{\mathrm{eff}}[O]=N$, and values above $N$ occur when the estimator is negatively correlated for that observable. Unlike $\ESS_w$, $N_{\mathrm{eff}}$ depends on the observable.

\subsection{Topological freezing and sector sampling}
\label{sec:u1_topology}

\begin{figure*}[!tp]
\centering\includegraphics[width=\textwidth]{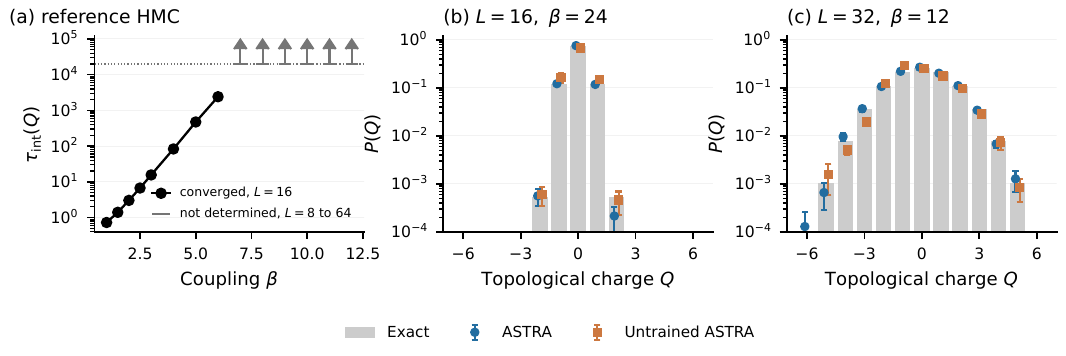}
\caption{In two-dimensional $U(1)$ theory, reference HMC freezes at large $\beta$, while ASTRA reproduces the exact charge distribution at couplings where reference HMC does not converge.
(a) Reference HMC estimates of $\tau_{\mathrm{int}}(Q)$ versus $\beta$. Each point uses $64$ chains of $20000$ trajectories. Points are converged estimates at $L=16$, and arrows mark couplings, for $L=8$ to $64$, at which $\tau_{\mathrm{int}}(Q)$ could not be determined within $2\times10^4$ trajectories. The dotted line marks $2\times10^4$. 
(b) and (c) Signed charge probabilities $P(Q)$ at $(L,\beta)=(16,24)$ and $(32,12)$. Gray bars show the exact finite-volume results. Points show the weighted means from independent runs of ASTRA and untrained ASTRA. Error bars indicate standard errors. Each setting has $16$ runs in both panels, and each run uses $N=1024$ particles. Untrained ASTRA deviates from the exact values by up to $4.8$ standard errors at $(32,12)$.
}
\label{fig:U1_topology}
\end{figure*}

We first look at topological freezing in reference HMC, the standard sampler run directly at the target coupling. Figure~\ref{fig:U1_topology}(a) shows the integrated autocorrelation time of the topological charge, $\tau_{\mathrm{int}}(Q)=\tfrac12+\sum_{n\ge1}\rho_Q(n)$, versus $\beta$ at several lattice sizes. Here $\rho_Q(n)$ is the normalized autocorrelation function of $Q$ at a separation of $n$ HMC trajectories. As $\beta$ grows, topological transitions become rare and $\tau_{\mathrm{int}}(Q)$ grows rapidly. For several parameters it cannot be determined within $2\times10^4$ HMC trajectories. Longer runs with $1024$ chains of $4\times10^5$ trajectories each show no topological transition for $L\geq16$ on the two lines of constant physics of Sec.~\ref{sec:u1_accuracy}, or at $(L,\beta)=(12,22.5)$.

We next test whether ASTRA reproduces the relative probabilities of the topological sectors. Figures~\ref{fig:U1_topology}(b) and \ref{fig:U1_topology}(c) compare the sampled charge distributions $P(Q)$ with the exact finite-volume results at $(L,\beta)=(16,24)$ and $(32,12)$. Full ASTRA is compared with untrained ASTRA.  Full ASTRA reproduces the exact sector probabilities at both points, within $2.7$ standard errors, with the largest deviation in a sector of probability $5\times10^{-4}$. Untrained ASTRA, with the same construction and number of particles, deviates by up to $4.8$ standard errors at $(32,12)$, so the learned residual improves the accuracy at a fixed particle number. The two settings differ in efficiency, which Sec.~\ref{sec:u1_ablation} quantifies.

Histories of the topological charge show the same contrast. Reference HMC chains at $(L,\beta)=(16,24)$ stay in their initial sectors over $2\times10^5$ trajectories, whereas ASTRA particles change sector in the early part of the path, while the coupling of the intermediate density is still weak, and settle by about $t=0.8$ [Fig.~\ref{fig:framework}(b) and (c) and Appendix~\ref{app:q_history}]. The path time $t$ of ASTRA measures progress along a path and is not Markov chain time, so the statistical performance of ASTRA is assessed through independent sampling runs and the accuracy of physical observables.

ASTRA thus reaches different topological sectors at couplings where reference HMC is frozen, and the exact sector probabilities test whether it weights them correctly.

\subsection{Accuracy of physical observables}
\label{sec:u1_accuracy}

We now test the topological susceptibility and Wilson loops against the exact finite-volume references of two-dimensional $U(1)$. These references measure both the statistical accuracy and the effects of the finite population.

\begin{figure*}[!tp]
\centering\includegraphics[width=\textwidth]{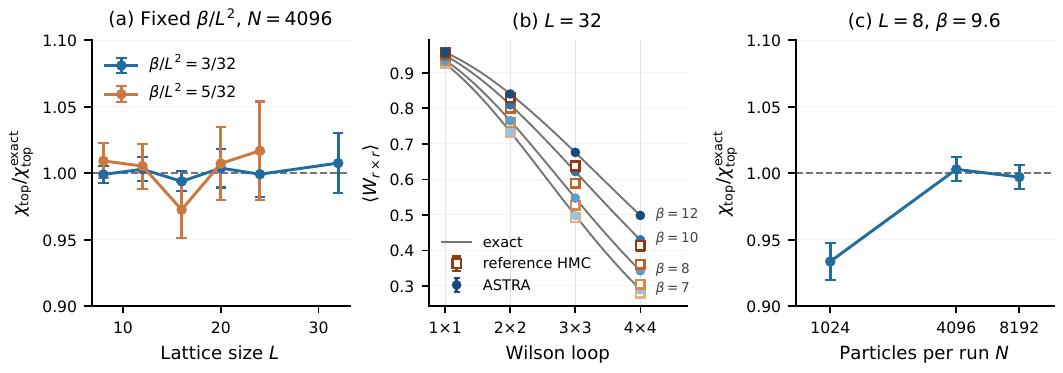}
\caption{Physical observables in two-dimensional $U(1)$ theory.
(a) The topological susceptibility relative to its exact value along two lines of constant physics, with $N=4096$ and $20$ independent runs per point.
(b) Wilson loops $\langle W_{r\times r}\rangle$ at $L=32$ for $64$ reference HMC chains with a hot start (open squares) and for ASTRA with the model trained at $L=8$ of Appendix~\ref{app:production} (filled circles, $N=1024$, $14$ to $20$ runs per coupling). Gray lines are the exact values, and darker colors mark larger $\beta$. (c) Dependence on the particle number at $(L,\beta)=(8,9.6)$, with ten independent runs per point.
Overbars denote means over independent runs or chains. Error bars show statistical errors, and the gray dashed line in (a) marks the exact reference.}
\label{fig:U1_physical_observables}
\end{figure*}

\begin{figure*}[!tp]
\centering\includegraphics[width=0.85\textwidth]{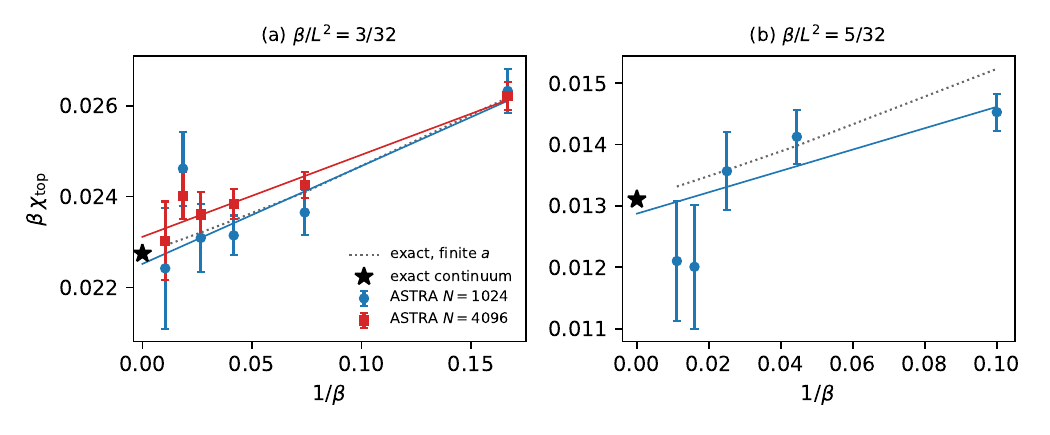}
\caption{Continuum extrapolation of the topological susceptibility along (a) $\beta/L^2=3/32$ and (b) $5/32$. Points are means over independent ASTRA runs with the model trained at $L=8$ ($18$ runs for $N=1024$, $20$ for $N=4096$). Solid lines are fits linear in $1/\beta$. Dotted lines show exact finite-volume values (exact, finite $a$) and stars the exact continuum values.}
\label{fig:U1_continuum}
\end{figure*}

Figure~\ref{fig:U1_physical_observables}(a) compares the topological susceptibility with its exact value along two lines of constant physics, $\beta/L^2=3/32$ and $5/32$. The first line has six points from $L=8$ to $32$, and the second has five from $L=8$ to $24$. Fixed $\beta/L^2$ keeps the physical area in units of the continuum coupling fixed, so the two lines test sampling as the lattice is refined. The figure shows runs with $N=4096$, and every point agrees with the exact value within $1.4$ standard errors. With $N=1024$ and $18$ runs per point, eight of the eleven points agree within $1.3$ standard errors and ten within two, and the largest deviations shrink when $N$ is increased. ASTRA therefore reproduces the topological fluctuations over this range of lattice sizes and couplings, including couplings where HMC is frozen.

Wilson loops provide complementary tests of the equilibrium measure beyond topological observables. Figure~\ref{fig:U1_physical_observables}(b) compares Wilson loops from $1\times1$ to $4\times4$ at $L=32$ with their exact values. At $\beta=12$, ASTRA with the model trained at $L=8$ gives
\begin{equation}
\frac{
\overline{W}_{4\times4}^{\mathrm{ASTRA}}
}{
\langle W_{4\times4}\rangle_{\mathrm{exact}}
}
=1.0004(9),
\end{equation}
from 20 independent sampling runs. At $\beta=7$, $8$, and $10$ the $4\times4$ loops lie within $0.3\%$ of the exact values. Reference HMC with a hot start keeps the topological charge of its random initial links, and its Wilson loops lie below the exact values by an amount that grows with the loop size and with $\beta$, down to $0.83$ of the exact value for the $4\times4$ loop at $\beta=12$. Observables measured at frozen topology are known to carry such a bias~\cite{Aoki2007FixedTopology}, and Sec.~S2 of the Supplemental Material relates it to the frozen charge of these chains. HMC with updates that shift the topological charge samples the topological susceptibility correctly along the line $\beta/L^2=0.094$ up to $L=32$, but becomes biased deeper in the frozen regime~\cite{Singha2026Sector}. The accuracy of ASTRA thus extends beyond the topological charge.

Figure~\ref{fig:U1_physical_observables}(c) shows how the topological susceptibility at $L=8$ and $\beta=9.6$ depends on the particle number. The ratios to the exact susceptibility are $0.934(14)$, $1.003(9)$, and $0.997(9)$ for $N=1024$, $4096$, and $8192$, respectively. The underestimate at $N=1024$ disappears at larger $N$. Topology-sensitive observables therefore need enough particles.
We measure accuracy by the standardized deviation $z_O=(\overline O-O_{\mathrm{ref}})/\mathrm{SE}(\overline O)$, the deviation of the mean over independent runs from the exact reference in units of its standard error. Across the three systems, $90\%$ of the $143$ comparisons satisfy $|z_O|<2$ (Appendix~\ref{app:pulls}).

Finally, we examine the continuum extrapolation of the topological susceptibility along the two lines of constant physics. Figure~\ref{fig:U1_continuum} shows $\beta\chi_{\mathrm{top}}$ as a function of $1/\beta$, which is proportional to $a^2$, since $\beta=1/(e^2a^2)$ in two dimensions with gauge coupling $e$ and lattice spacing $a$. At fixed $\beta/L^2$, the continuum topological charge distribution is~\cite{Bonati:2019ylr}
\begin{equation}
P(Q)\propto
\exp\left[
-2\pi^2\frac{\beta}{L^2}Q^2
\right].
\label{eq:u1_continuum_charge}
\end{equation}
This gives continuum reference values $\beta\chi_{\mathrm{top}}=0.02275$ and $0.01311$ for the two lines, respectively. Linear extrapolations in $1/\beta$ yield $0.02252(41)$ and $0.02256(17)$ along the first line for $N=1024$ and $4096$, respectively, and $0.01287(49)$ and $0.01292(23)$ along the second line. All four agree with the continuum values within $1.1$ standard errors.

With enough particles, ASTRA thus reproduces the exact topological susceptibility and Wilson loops of two-dimensional $U(1)$, and its continuum extrapolations agree with the exact values.

\subsection{Learned residual, resampling, and HMC moves}
\label{sec:u1_ablation}
\label{sec:population}

This subsection asks what the learned residual and each SMC component add.

Ablations with the model held fixed at $L=16$ and $\beta=24$ (Appendix~\ref{app:ablation}) show the role of each sampling component. Resampling and HMC moves reduce the spread of single runs by a factor of three to four and a half compared with no HMC moves or with independent paths. For untrained ASTRA, the spread is about four times larger and resampling is about four times more frequent. In the early part of the path, HMC moves after resampling change the topological charge of most particles. After about $t=0.8$ the charge of a particle almost never changes, and the moves only separate duplicated particles within each sector (Appendix~\ref{app:q_history}).

Resampling also concentrates the population on a few lines of descent. We therefore record the initial ancestor of every particle and follow its topological charge. Full ASTRA keeps a median of $30$ to $64$ distinct ancestors in each occupied topological sector, while untrained ASTRA keeps about $7$ to $14$. The two settings use the same construction and path length and differ only in the learned residual. Appendix~\ref{app:u1_production} gives these counts.

The final weights vary mainly within each $|Q|$ sector. The part of the weight variance between sectors is several hundred times smaller than the part within sectors (Appendix~\ref{app:resampling}).

The resampling threshold balances two effects. Rare resampling leaves the weights on a few paths, and frequent resampling creates duplicates that the HMC moves must separate. Between $\ESS_w/N=0.3$ and $0.9$ all estimates of $\chi_{\rm top}$ agree with the exact value, and $N_{\rm eff}[\chi_{\rm top}]$ varies between $209$ and $417$ without a trend (Appendix~\ref{app:u1_production}). Annealed SMC schedules have been optimized by treating particle number, annealing steps, and resampling together~\cite{Syed2026OASMC}.

The learned residual thus reduces the concentration of weights and the number of resampling steps, and the HMC moves keep the particles diverse.

Along the first line, the cost of this accuracy grows as the lattice is refined. From $L=8$ to $L=32$, $N_{\mathrm{eff}}[\chi_{\rm top}]/N$ falls from $0.45$ to $0.053$, and the number of initial ancestors with surviving descendants falls from $699$ to $45$ (Appendix~\ref{app:run-stats}). For $L\ge16$ on both lines this number stays close to $1.6\,N/n_{\rm rs}$, where $n_{\rm rs}$ is the number of resampling steps per run, and $n_{\rm rs}$ grows slightly faster than the volume (Sec.~S4 of the Supplemental Material). The number of surviving ancestors therefore falls about inversely with the volume.

At $L=32$ and $\beta=12$, the number of surviving ancestors with $N=4096$ is about four times larger than with $N=1024$, and the standard deviation of a single run is about half as large (Appendix~\ref{app:u1_production}).

\section{\texorpdfstring{ASTRA for two-dimensional $SU(3)$ gauge theory}{ASTRA for two-dimensional SU(3) gauge theory}}
\label{sec:su3}\label{sec:gauge}
We now extend ASTRA from the Abelian $U(1)$ gauge theory to two-dimensional non-Abelian $SU(3)$ gauge theory. Unlike $U(1)$, whose gauge variables are characterized by a single compact angle, $SU(3)$ is a non-Abelian Lie group with eight generators. Its link variables are $3\times3$ special unitary matrices satisfying
\begin{equation}
U_{z,\mu}\in SU(3),\qquad
U_{z,\mu}^{\dagger}U_{z,\mu}=\mathbb{I},
\qquad
\det U_{z,\mu}=1.
\label{eq:su3_links}
\end{equation}
We consider a periodic two-dimensional square lattice with $V=L^2$ sites and $2V$ oriented links. The elementary plaquette variable is defined as
\begin{equation}
U_{\square}(z)
=
U_{z,0}
U_{z+\hat{0},1}
U_{z+\hat{1},0}^{\dagger}
U_{z,1}^{\dagger},
\label{eq:su3_plaquette}
\end{equation} 
where the ordering of the link matrices is essential because the group multiplication is noncommutative. The Wilson gauge action is given by
\begin{equation}
S_{\beta}(U)
=
-\frac{\beta}{3}
\sum_z
\operatorname{Re}\operatorname{Tr}
U_{\square}(z),
\label{eq:su3_action}
\end{equation}
where $\beta$ denotes the lattice coupling. Under a local gauge transformation $\Omega_z\in SU(3)$, the link variables transform as
\begin{equation}
U_{z,\mu}
\longrightarrow
\Omega_z U_{z,\mu}
\Omega_{z+\hat{\mu}}^{\dagger},
\label{eq:su3_gauge_transformation}
\end{equation}
leaving the Wilson action invariant. Appendix~\ref{app:su3-construction} gives the noise process on the group, the physics term, and the path weights.
Two-dimensional pure $SU(3)$ gauge theory has no local propagating degrees of freedom.  Two-dimensional pure gauge theory also admits specialized sampling methods that exploit the plaquette structure and global holonomy, including conditional sampling and corner reweighting~\cite{Komijani2026Plaquette}. Character expansions give exact finite-volume results for gauge-invariant observables. We therefore use this theory to test ASTRA on non-Abelian gauge fields, through the accuracy of Wilson loops, the transfer across volumes, and the sampling efficiency.

\subsection{Wilson loops and volume generalization}
\label{sec:su3_wilson}

A model conditioned on couplings from $\beta=6$ to $16$ is trained only at $L=8$ and applied to $L=16$ and $32$ without retraining. Each sampling run uses $N=256$ particles and $T=1024$ path steps, with 64 independent runs for each parameter set.

Wilson loops provide a direct test of the generated gauge-field configurations. For a rectangular loop $\mathcal C_{r,s}$, we define the normalized Wilson-loop observable as
\begin{equation}
W_{r\times s}
=
\frac{1}{3V}
\sum_z
\operatorname{Re}\operatorname{Tr}
\left[
\prod_{\ell\in\mathcal C_{r,s}(z)}
U_\ell
\right],
\label{eq:su3_wilson_loop}
\end{equation}
where the link matrices are ordered along the closed contour, with inverse links represented by Hermitian conjugation. Appendix~\ref{app:su3-reference} gives the exact values.

\begin{figure*}[!tp]
\centering\includegraphics[width=\textwidth]{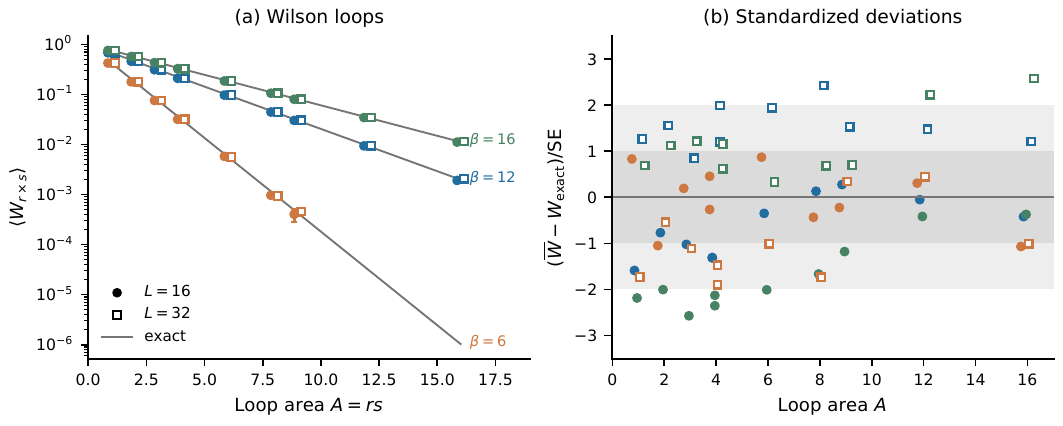}
\caption{Wilson loops in two-dimensional $SU(3)$ gauge theory from a model trained at $L=8$. (a) $\langle W_{r\times s}\rangle$ versus loop area at $L=16$ (filled circles) and $L=32$ (open squares). Gray lines are exact values. (b) Standardized deviations from the exact values, with colors and markers as in (a). Gray bands mark one and two standard errors. At $\beta=6$ the loops of area $12$ and $16$ are smaller than three standard errors and are omitted from (a), and their deviations are shown in (b).}
\label{fig:su3-wilson}
\end{figure*}

Figure~\ref{fig:su3-wilson}(a) compares ASTRA estimates of rectangular Wilson loops with areas up to $A=16$ at $\beta=6$, $12$, and $16$. The results on $L=16$ and $32$ follow the exact area law over these areas. The standardized deviations in Fig.~\ref{fig:su3-wilson}(b) lie within $2.6$ standard errors. Within each $(L,\beta)$ they shift together, because all loops come from the same runs.

As a representative example, at $\beta=12$ and $L=32$, ASTRA gives a mean plaquette of $0.677727(44)$ and a $1\times2$ Wilson loop of $0.45935(7)$, compared with the exact values $0.677672$ and $0.459239$, respectively. Additional numerical comparisons at $\beta=6$ and $16$ are reported in Appendix~\ref{app:extended}. The agreement shows that the model trained at $L=8$ reproduces gauge-invariant observables on larger $SU(3)$ lattices without retraining.

Larger volumes need more frequent resampling. At $\beta=16$ the number of resampling steps per run grows about tenfold from $L=8$ to $L=32$ (Appendix~\ref{app:extended}). Diffusion models for $SU(N)$ have studied score learning for $SU(3)$ in two and four dimensions, including the effect of molecular dynamics updates on sample quality and cost~\cite{Komijani2026SUN}.

\section{Conclusions}
\label{sec:conclusion}
We have introduced ASTRA, which transports a population of configurations from a tractable distribution to the target in a finite number of steps and corrects it by SMC. The corrected configurations train the model of the next round, so training needs no precomputed equilibrium data. The learned residual is local and conditioned on the coupling. One model per gauge group, trained at $L=8$, is therefore used on lattices up to $L=32$ without retraining, and the SMC correction keeps the estimates accurate there.

In an asymmetric double well, all five ASTRA models trained from scratch keep both wells, including the less probable one. Three of five models trained by reverse KL in path space lose one. ASTRA keeps the wells because the correction keeps the relative weight of each well that the corrected configurations reach. In two-dimensional compact $U(1)$ gauge theory, ASTRA reproduces the exact topological susceptibility up to $\beta=96$, where reference HMC does not change the topological charge. Its continuum extrapolation along two lines of constant physics agrees with the exact value, and its Wilson loops agree with the exact results. In two-dimensional $SU(3)$ gauge theory, ASTRA reproduces the exact Wilson loops from the plaquette up to area $16$. The construction thus carries over to non-Abelian gauge fields.

A model that has lost a mode does not regain it from its own corrected configurations. In the double well, a small fraction of training paths drawn with the Gaussian drift alone restores it. In our runs the number of surviving initial ancestors falls about inversely with the lattice volume. The effective sample size of $\chi_{\rm top}$ per particle falls about eightfold from $L=8$ to $L=32$. This sets the cost at larger volumes and is the main quantity to improve. Future work will reduce the growth of the required particle number with the volume. It will also apply ASTRA to four-dimensional $SU(3)$ gauge theory and to two-dimensional sigma models, such as the $O(3)$ and $CP^{N-1}$ models, which also have topological sectors. Dynamical fermions can enter through pseudofermion fields transported together with the gauge links.

Two features make ASTRA useful where Markov chains freeze. First, the SMC correction makes the estimates consistent as the particle number grows. This holds for any model whose intermediate densities cover the target, and the model only sets how many particles a given accuracy needs. Second, sampling runs are independent and start from fresh populations. Errors therefore follow from the spread between runs without autocorrelation analysis, and the runs parallelize trivially. Finally, the construction needs only the action and its gradient. It therefore applies to other Boltzmann distributions in statistical physics, as the double well and $\phi^4$ theory in the Supplemental Material illustrate.
\begin{acknowledgments}
We acknowledge computing time on the HIMster III cluster at the Helmholtz Institute Mainz and on the computing cluster of the Institute for Quantitative and Computational Biosciences (IQCB) at Johannes Gutenberg University Mainz. Claude Code (Anthropic, Claude Opus 5.5) was used to polish the language of the manuscript, and the authors take full responsibility for its content.

\end{acknowledgments}

\section*{Data and code availability}
The sampling code, the trained models, the data for each run, and representative logs are available at \url{https://github.com/qxxmax/astralattice}.
Appendix~\ref{app:replicates} gives the independent runs and training seeds.

\FloatBarrier
\appendix

\section{Additional results for the asymmetric double well}
\label{app:doublewell_results}\label{app:status}

For numerical stability, each drift step in the double well is smoothly capped so that it moves each coordinate by less than one unit. The training runs of Fig.~\ref{fig:double_well} also resample at intermediate times, with HMC moves on the intermediate density.

\subsection{Mode retention and reweighting strategies}
\label{app:dw_reweight}\label{app:reweight}

\begin{figure*}[t]
\centering
\includegraphics[width=0.92\textwidth]{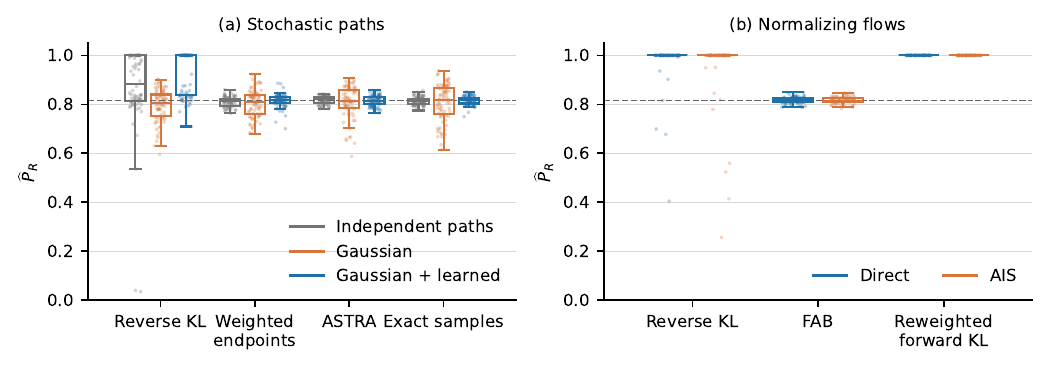}
\caption{Double-well sampling estimates $\widehat P_R$ across training objectives and sampling schemes, five trained models with $16$ runs each per setting. (a) Training by reverse KL in path space, on weighted endpoints of independent paths, by ASTRA, and on exact samples, each evaluated with independent paths, the Gaussian intermediate density, which contains no information about the target, and the model's own intermediate density, labeled Gaussian + learned. (b) Normalizing flows trained by reverse KL, by flow annealed importance sampling bootstrap (FAB)~\cite{Midgley2023}, and on their own reweighted samples, each evaluated directly and with annealed importance sampling (AIS) in $32$ stages. The dashed line is the exact $P_R$. Boxes show the quartiles and the median, whiskers the range without outliers, and points the single runs. Table~\ref{tab:dw_weights} gives the retention counts of the five strategies.}
\label{fig:app_dw_summary}
\end{figure*}

For each of five training strategies, we train ten independent models on configurations covering both wells. A model counts as retaining both wells only when all 16 sampling runs visit both. Table~\ref{tab:dw_weights} summarizes the retention counts of the five strategies. Figure~\ref{fig:app_dw_summary} shows the individual sampling estimates. The strategies differ in how the corrected weights enter training, as defined in Sec.~S3 of the Supplemental Material. Fully weighted regression and endpoint resampling with HMC retain both modes more often than unweighted force regression. Their mean right-well probabilities are $0.8157(13)$ and $0.8197(16)$, respectively, compared with the exact $P_R=0.8153$.

\begin{table}[t]
\centering
\caption{Mode retention in the asymmetric double well. Each strategy is evaluated with ten independently trained models, and each model has 16 sampling runs. The criterion requires both wells to be present in every sampling run. The ESS-adaptive rule uses $\kappa=0.8$ (Sec.~S3 of the Supplemental Material). These models omit the physics term, use $\sigma_{\max}=1.5$, and are sampled with the learned intermediate density.}
\label{tab:dw_weights}
\resizebox{\columnwidth}{!}{%
\begin{tabular}{lc}
\toprule
Training strategy & Models retaining both wells \\
\midrule
Unweighted force regression ($\eta=0$) & $1/10$ \\
ESS-adaptive weights & $1/10$ \\
Linearly increasing $\eta$ & $4/10$ \\
Fully weighted regression ($\eta=1$) & $9/10$ \\
Endpoint resampling + HMC & $8/10$ \\
\bottomrule
\end{tabular}%
}
\end{table}

\subsection{Recovery of a lost mode}
\label{app:dw_recovery}

\begin{figure*}[t]
\centering
\includegraphics[width=\textwidth]{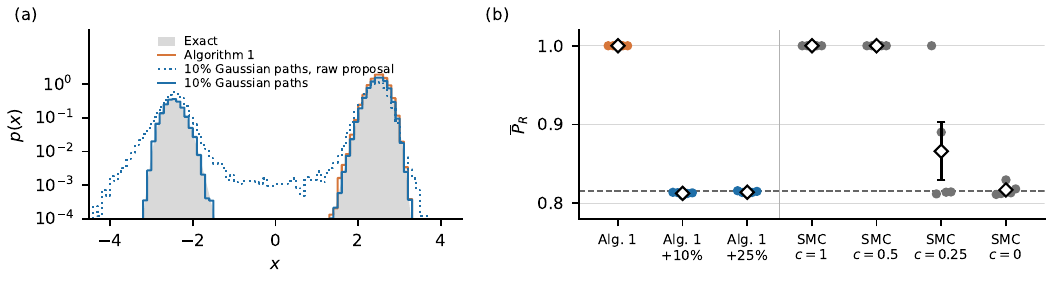}
\caption{Recovery of a lost mode in the double well, for five models pretrained on right-well samples only and sampled with their own intermediate density. (a) Distribution of the final $x$, pooled over five models with $16$ runs of $N=1024$ each, for training with Algorithm~\ref{alg:training} (orange) and with $10\%$ of the training paths following the Gaussian drift (blue). The dotted line shows the generated configurations of the recovered models, the endpoints of their independent paths before weighting and resampling, and gray shows the exact distribution. (b) $\overline P_R$ of each model (circles) and the mean with its standard error (diamonds), for Algorithm~\ref{alg:training} alone and with $10\%$ or $25\%$ Gaussian paths, and for runs that resample at intermediate times with the learned residual scaled by $c$ in the intermediate density. The dashed line is the exact $P_R$.}
\label{fig:dw_recovery}
\end{figure*}

Five models are pretrained for $2000$ steps by force regression on a buffer filled with exact samples from the right well only, so each starts mode collapsed. They then run $20$ training rounds of Algorithm~\ref{alg:training}, with $1024$ independent paths of $128$ steps per round, endpoint resampling, and HMC moves targeting the double well. Each model is evaluated with $16$ sampling runs of $N=1024$ particles that use its own intermediate density (Fig.~\ref{fig:dw_recovery}).

With Algorithm~\ref{alg:training} alone, the paths of a collapsed model never reach the left well, and all five models stay collapsed with $\overline P_R=1$. We therefore draw a fraction $\alpha$ of the training paths with the drift of the Gaussian density alone, without the learned residual, and the rest with the learned drift. This is a defensive mixture in the sense of importance sampling~\mbox{\cite{Hesterberg1995}}. Each path is weighted with the mixture $(1-\alpha)\mathcal Q_\theta+\alpha\mathcal Q_0$ of the two path densities, where $\mathcal Q_0$ is the path density with the Gaussian drift, so the weights stay exact. With $\alpha=0.1$ all five models regain the left well within the first two rounds, and sampling gives $\overline P_R=0.8126(7)$ at $\ESS_w/N=0.71$. With $\alpha=0.25$ it gives $0.8141(7)$. Both lie slightly below the exact $0.8153$. During training, $\ESS_w$ of the endpoints is about $0.2N$.

Figure~\ref{fig:dw_recovery}(b) also shows runs that resample at intermediate times, as in Fig.~\ref{fig:double_well}, with the learned residual scaled by a factor $c$ in the intermediate density, $\gamma_t(x)\propto\exp[-|x|^2/(2\sigma^2(t)+2)+c\,\mathcal E_\theta(x,t)]$. Sampling afterward always uses $c=1$. With the model's own density, $c=1$, and with $c=0.5$ all five models stay collapsed, because $e^{c\mathcal E_\theta}$ still multiplies the density in the left well. With $c=0.25$ three of five recover. The Gaussian alone, $c=0$, recovers all five with $\overline P_R=0.8169(34)$, but $\ESS_w$ at the endpoint of these runs is only about $0.02N$.

\subsection{Effect of sample reuse}
\label{app:dw_replay}

\begin{figure*}[t]
\centering
\includegraphics[width=\textwidth]{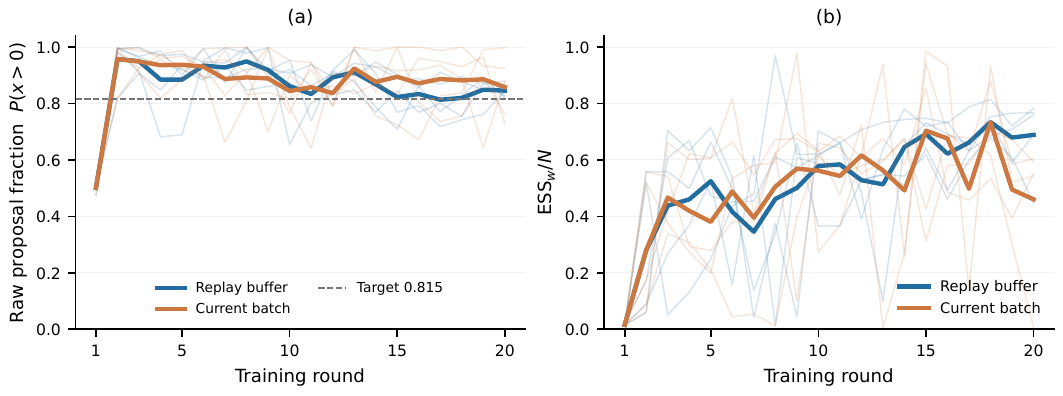}
\caption{Replay-buffer comparison for the double well. (a) Right-well fraction of the generated configurations, with the dashed line at the exact $P_R$. (b) ESS of the endpoint weights over training rounds. Thin lines show the five training seeds, and thick lines their mean.}
\label{fig:app_dw_replay}
\end{figure*}

Replay-buffer training and current-batch-only training are compared with five training seeds, 20 rounds, and 100 updates per round. The buffer accumulates 1024 corrected configurations per round (20480 by the final round), whereas the current-batch baseline retains 1024. Both settings omit the physics term and use $\sigma_{\max}=1.5$. As Fig.~\ref{fig:app_dw_replay} shows, the standard deviation across models of the right-well fraction of the generated configurations, the endpoints of independent paths before weighting and resampling, at the final round decreases from $0.112$ without replay to $0.024$ with replay. For the configurations used in training, which in the replay setting include all stored rounds, it decreases from $0.115$ to $0.016$.

\section{Additional results for two-dimensional compact $U(1)$}
\label{app:u1_results}

\subsection{Physics-term comparison}
\label{app:u1_physics_ablation}\label{app:no-analytic}

Table~\ref{tab:u1_physics} compares models whose network outputs the drift directly, trained with and without the physics term on an $L=8$ lattice. Each entry is the mean over five trained models, evaluated using $N=16384$ independent paths of length $T=4096$. With the physics term $\ESS_w/N$ is higher at every tested coupling, and both variants give $\chi_{\rm top}$ close to the exact value.

\begin{table}[t]
\centering
\caption{Two-dimensional compact $U(1)$, $L=8$: training with and without the physics term. Parentheses give standard errors over five trained models. $\mathrm{ESS}_w$ is the importance-weight ESS of independent paths. These models output the force directly and are trained at $\beta=6$ to $12$ for the rows $\beta\le12$ and at $\beta=16$ to $32$ for the rows $\beta\ge16$.}
\label{tab:u1_physics}\label{tab:physics-term}
\resizebox{\columnwidth}{!}{%
\begin{tabular}{rcccc}
\toprule
& \multicolumn{2}{c}{With physics term} & \multicolumn{2}{c}{Without physics term} \\
$\beta$ & $\mathrm{ESS}_w/N$ & $\chi_{\rm top}/\chi_{\rm top}^{\rm exact}$ & $\mathrm{ESS}_w/N$ & $\chi_{\rm top}/\chi_{\rm top}^{\rm exact}$ \\
\midrule
8 & 0.346(16) & 1.000(11) & 0.312(7) & 0.997(14) \\
10 & 0.271(23) & 1.007(16) & 0.243(4) & 1.010(4) \\
12 & 0.228(13) & 0.997(14) & 0.195(4) & 0.988(13) \\
16 & 0.129(8) & 0.983(13) & 0.094(5) & 0.992(20) \\
24 & 0.088(16) & 1.018(27) & 0.079(5) & 1.024(14) \\
32 & 0.090(6) & 1.000(21) & 0.078(6) & 1.036(45) \\
\bottomrule
\end{tabular}%
}
\end{table}

\begin{figure}[t]
\centering
\includegraphics[width=\columnwidth]{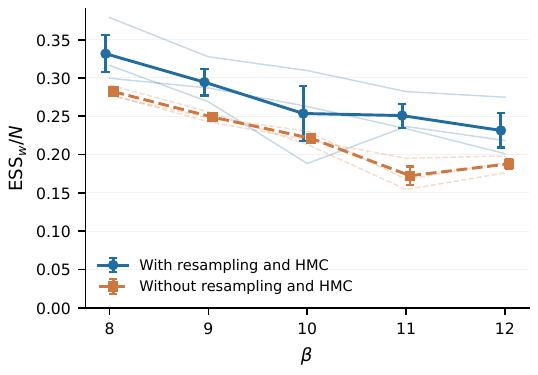}
\caption{$\ESS_w/N$ of independent paths in $U(1)$ at $L=8$ for training with and without endpoint resampling and HMC moves. Markers show the mean over three trained models with standard errors, and thin lines show the single models.}
\label{fig:app_u1_training}
\end{figure}

For $L=8$ and $\beta=8$ to $12$, the tested training strategies give $\mathrm{ESS}_w/N=0.23$ to $0.35$ with endpoint resampling and HMC, $0.19$ to $0.30$ with ESS-adaptive weights, $0.18$ to $0.27$ with unweighted force regression, and $0.010$ to $0.018$ with fully weighted regression. Figure~\ref{fig:app_u1_training} compares training with and without endpoint resampling and HMC moves.

\subsection{Sampling ablations and particle number}
\label{app:u1_production}\label{app:ablation}

\begin{figure*}[t]
\centering
\includegraphics[width=0.95\textwidth]{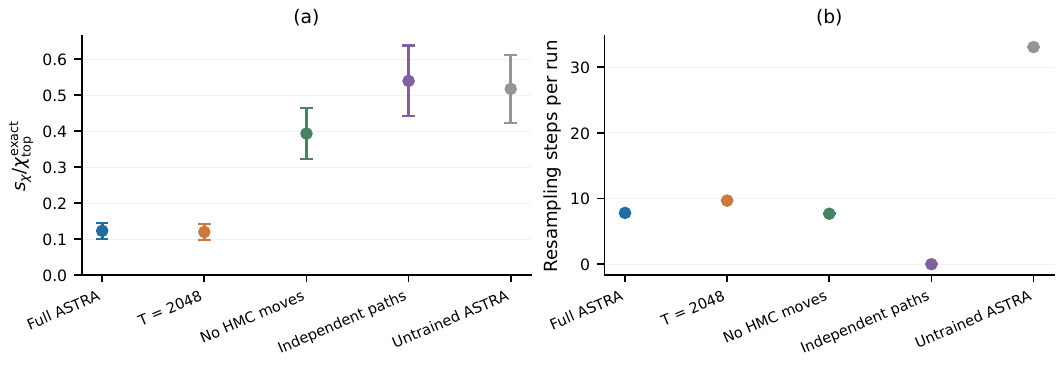}
\caption{Sampling ablations at $(L,\beta)=(16,24)$ with $N=1024$ and $16$ runs per setting. Resampling with HMC moves reduces the spread of single runs, and the learned residual reduces the number of resampling steps. (a) Standard deviation of single-run estimates of $\chi_{\rm top}$ relative to the exact value, with its uncertainty. (b) Resampling steps per run. Table~\ref{tab:u1_prod} gives the means and the AIS settings.}
\label{fig:app_u1_ablation}
\end{figure*}

Table~\ref{tab:u1_prod} and Fig.~\ref{fig:app_u1_ablation} remove one sampling component at a time at $(L,\beta)=(16,24)$, with $N=1024$ and $16$ independent runs per setting. Full ASTRA has a smaller spread between runs than the settings without HMC moves, with independent paths, or for untrained ASTRA.

\begin{table}[t]
\centering
\caption{Sampling ablations at $(L,\beta)=(16,24)$, $N=1024$. The standard deviation $s_\chi$ of the estimates of single runs measures the spread between runs, with $\chi_{\rm top}^{\rm exact}=9.783\times10^{-4}$. Except where specified, learned paths use $T=4096$. All $53$ runs with this model at this point and $N=1024$ give $0.963(14)$. AIS denotes annealed importance sampling with resampling and one HMC trajectory at every one of its $K$ steps, without learning.}
\label{tab:u1_prod}\label{tab:U1_physical_observables}
\resizebox{\columnwidth}{!}{%
\begin{tabular}{lccc}
\toprule
Setting & $\chi_{\rm top}/\chi_{\rm top}^{\rm exact}$ & $s_\chi/\chi_{\rm top}^{\rm exact}$ & Resampling steps/run \\
\midrule
Full ASTRA & 0.92(3) & 0.12 & 7.8 \\
$T=2048$ & 1.04(3) & 0.12 & 9.7 \\
No HMC moves & 1.02(10) & 0.39 & 7.7 \\
Independent paths & 0.92(14) & 0.54 & 0 \\
Untrained ASTRA & 1.19(13) & 0.52 & 33.1 \\
AIS, $K=256$ & 1.35(28) & 1.11 & 36.3 \\
AIS, $K=16384$ & 1.00(2) & 0.06 & 2.1 \\
\bottomrule
\end{tabular}%
}
\end{table}

At the end of the path, full ASTRA keeps a median of $64$ distinct initial ancestors per occupied sector at $(16,24)$ and $30$ at $(32,12)$, counting sectors with more than $1\%$ of the weight. Untrained ASTRA keeps $14$ and $7$.

The ESS thresholds $0.3$, $0.5$, $0.7$, and $0.9$ give about $5$, $8$, $13$, and $33$ resampling steps per run and $N_{\mathrm{eff}}[\chi_{\rm top}]=209$, $309$, $417$, and $222$ ($16$ runs each), whose $68\%$ intervals overlap. Section~S4 of the Supplemental Material gives scans of the path length and of the number of HMC moves.

At $L=32$ and $\beta=12$, increasing $N$ from $1024$ to $4096$ ($8$ runs each) raises the number of surviving initial ancestors $N_{\rm anc}$ (Appendix~\ref{app:ancestors}) from $107$ to $439$ and $N_{\mathrm{eff}}[\chi_{\rm top}]$ from $124$ to $403$. The ratios to the exact $\chi_{\rm top}$ are $0.992(45)$ and $1.000(25)$, and the standard deviation of a single run falls from $0.13$ to $0.07$ of the exact value.

\subsection{Histories of the topological charge}
\label{app:q_history}

\begin{figure*}[!tp]
\centering\includegraphics[width=\textwidth]{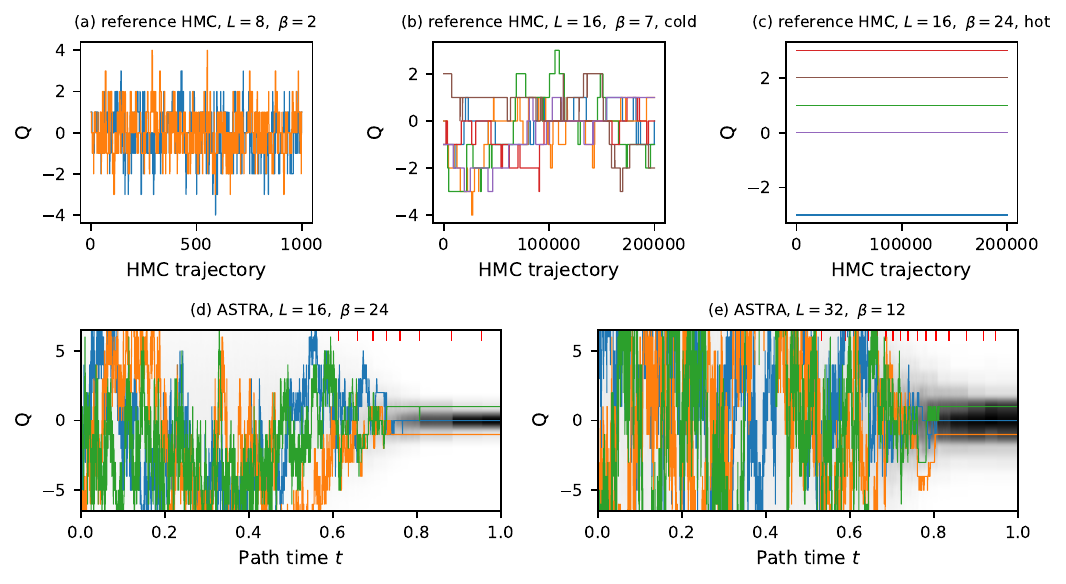}
\caption{Histories of the topological charge. (a) to (c) Representative reference HMC chains from separate runs of $16$ chains, longer than those of Fig.~\ref{fig:U1_topology}(a), shown versus HMC trajectory number for $(L,\beta)=(8,2)$, $(16,7)$ (cold start, all links set to one), and $(16,24)$ (hot start, random links). Panel (a) shows two chains over the first 1000 trajectories, and panels (b) and (c) show six chains over 200000 trajectories, and chains in the same sector overlap. (d) and (e) Particle histories along the path for ASTRA at $(16,24)$ and $(32,12)$. The gray shading indicates occupied sectors, colored lines show the histories of three final particles with $Q=0$, $-1$, and $1$, traced back through the resampling steps, and red ticks at the top edge mark resampling followed by HMC moves.}
\label{fig:U1_topological-history}
\end{figure*}

Figure~\ref{fig:U1_topological-history} shows histories of the topological charge in reference HMC and ASTRA. Panels (a) to (c) show reference HMC chains at $(L,\beta)=(8,2)$, $(16,7)$, and $(16,24)$. At small $\beta$ the chains change their charge often, while at larger $\beta$ they stay in one sector for long periods. At $(16,24)$ they stay frozen over all displayed trajectories.

In panels (d) and (e), ASTRA particles visit several sectors along the path at $(16,24)$ and $(32,12)$. Propagation steps, resampling, and HMC moves let the particles reach different sectors.

HMC moves after resampling change the topological charge of $53\%$ of the particles at $(L,\beta)=(16,24)$ and of $66\%$ at $(32,12)$, averaged over the path. These changes happen early on the path. In the runs with $N=4096$ on both lines of constant physics, the fraction of particles whose charge changes within $16$ path steps falls below $1\%$ at $t=0.81$ to $0.83$ for every $L$ from $8$ to $32$ (median over $20$ runs). After that time the sector weights change mainly through resampling.

\section{Additional results for two-dimensional $SU(3)$}
\label{app:su3_results}\label{app:extended}

\subsection{Exact-reference observables and volume transfer}
\label{app:su3_transfer}

A model trained only at $L=8$ for $\beta=6$ to $16$ is evaluated at $L=16$ and $32$ without retraining. Each point uses $N=256$ particles, $T=1024$ path steps, and 64 independent sampling runs. Table~\ref{tab:su3_transfer} collects the plaquette and $1\times2$ Wilson-loop values. The references follow from Appendix~\ref{app:su3-reference}, and Fig.~\ref{fig:su3-wilson} shows the Wilson loops.

\begin{table}[t]
\centering
\caption{Two-dimensional $SU(3)$ volume transfer. Parentheses denote standard errors of the mean over 64 independent sampling runs.}
\label{tab:su3_transfer}\label{tab:su3-extra}
\begin{tabular}{rrcc}
\toprule
$L$ & $\beta$ & Plaquette & $W_{1\times2}$ \\
\midrule
16 & 6 & 0.42265(14) & 0.17834(19) \\
16 & 12 & 0.67752(9) & 0.45912(16) \\
16 & 16 & 0.75522(7) & 0.57031(14) \\
32 & 6 & 0.422414(68) & 0.17849(9) \\
32 & 12 & 0.677727(44) & 0.45935(7) \\
32 & 16 & 0.755394(35) & 0.570655(63) \\
\midrule
Exact & 6 & 0.422532 & 0.178533 \\
Exact & 12 & 0.677672 & 0.459239 \\
Exact & 16 & 0.755370 & 0.570584 \\
\bottomrule
\end{tabular}
\end{table}

\subsection{Volume-dependent particle diagnostics}
\label{app:su3_efficiency}

At $\beta=16$ a sampling run has on average about $7$, $21$, and $67$ resampling steps at $L=8$, $16$, and $32$. About $57$, $21$, and $6$ distinct initial ancestors survive to the endpoint.

At $L=8$, $N_{\rm eff}[W_{1\times1}]/N$ for the trained model is $0.75(19)$, $1.11(28)$, and $1.17(30)$ for $\beta=6$, $12$, and $16$. Untrained ASTRA gives corresponding point estimates $1.18$, $1.29$, and $1.17$ (each from eight independent runs). At this volume untrained ASTRA thus gives the same plaquette efficiency as the trained model.

\section{\texorpdfstring{Physics term, path weights, and the construction for $SU(3)$}{Physics term, path weights, and the construction for SU(3)}}
\label{app:compact}
This appendix gives the drift of Sec.~\ref{sec:u1} in link coordinates, which is $-\beta_{\mathrm{eff}}(t)\sin\vartheta$ plus the gradient of the learned residual [Eq.~\eqref{eq:score}]. It also derives the identity for path weights on compact link angles, with its conditions. Appendix~\ref{app:weakfield} derives $\beta_{\rm eff}(t)$, and Appendix~\ref{app:su3-construction} gives the corresponding construction for $SU(3)$.

Under the support and integrability conditions, integrating the normalized backward kernels successively gives $\E_{\mathcal Q_\theta}[w_{\mathrm{path}}O(x_T)]=Z_\beta\E_{p_\beta}[O]$ for an integrable observable $O$. Here $\mathcal Q_\theta$ is the forward path measure, $q_{\mathrm B}=\prod_kb_k$, $\mathcal P_{\mathrm B}=p_\beta(x_T)q_{\mathrm B}$ is the backward path measure, and $w_{\mathrm{path}}/Z_\beta=\mathcal P_{\mathrm B}/\mathcal Q_\theta$. For independent paths, $\ESS_w/N$ is therefore related to the divergence in path space $\chi^2(\mathcal P_{\mathrm B}\Vert\mathcal Q_\theta)$, and $\ESS_w/N\to1/(1+\chi^2)$ for large $N$~\mbox{\cite{Agapiou2017}}.

With $\vartheta=Cx$, the scalar potential defines the drift
\begin{equation}
 s_\theta=-\nabla_xS_\theta
 =C^{\mathsf T}\left[-\beta_{\mathrm{eff}}(t)\sin\vartheta
                  +\nabla_\vartheta\mathcal E_\theta\right].
 \label{eq:score}
\end{equation}
The drift, weights, and HMC moves use the same intermediate density,
\begin{align}
 \log\gamma_k(x)&=-S_\theta(x,t_k;\beta)\nonumber\\
 &=\beta_{\mathrm{eff}}(t_k)\sum_z\cos\vartheta_z(x)\nonumber\\
 &\quad+\mathcal E_\theta(\vartheta(x),t_k;\beta).
 \label{eq:learned-bridge}
\end{align}
On a noncompact space, $0<\int\gamma_k(x)\dd x<\infty$ is required.
A shift $S_\theta(\cdot,t_k)\mapsto S_\theta(\cdot,t_k)+c_k$ cancels between the initial, incremental, and endpoint weights, since
\begin{equation}
 e^{-c_0}\prod_{k=0}^{T-1}e^{-(c_{k+1}-c_k)}e^{c_T}=1.
 \label{eq:bridge-shift}
\end{equation}

The Gaussian update in Sec.~\ref{sec:u1} can be represented as a joint kernel on angles and integer lifts.
This preserves periodicity and gives path weights for the discrete update.
For convenience, this appendix uses Lebesgue measure on $\mathcal C=[-\pi,\pi)^d$.
The relation to the main text is $\dd\mu(x)=\dd x/(2\pi)^d$.
All initial densities and transition kernels in this appendix are relative to Lebesgue measure.

Let $x\in\mathcal C$ be the angle configuration before an update and $x'\in\mathcal C$ the angles after it.
Write the generated unwrapped coordinate as $u=x'+2\pi n$, with $n\in\Z^d$.
Suppressing time indices, define
\begin{align}
 F(x',n\mid x)&=f(x'+2\pi n\mid x),\\
 B(x,n\mid x')&=b(x\mid x'+2\pi n).
\end{align}
The forward Gaussian on $\R^d$ satisfies
$\sum_n\int_{\mathcal C}F(x',n\mid x)\dd x'=1$.
The drift-free backward Gaussian is covariant under simultaneous shifts, meaning that
\begin{equation}
 b(x+2\pi m\mid u+2\pi m)=b(x\mid u).
\end{equation}
Hence
\begin{align}
 \sum_n\int_{\mathcal C}B(x,n\mid x')\dd x
 &=\sum_n\int_{\mathcal C}b(x-2\pi n\mid x')\dd x\nonumber\\
 &=\int_{\R^d}b(v\mid x')\dd v=1.
\end{align}
The backward kernel is normalized on the joint space of angles and integer lifts.

On this space, the incremental weight
\begin{equation}
 G(x,x',n)=\frac{\gamma_{k+1}(x')B(x,n\mid x')}
                   {\gamma_k(x)F(x',n\mid x)}
\end{equation}
satisfies the change-of-measure identity
\begin{align}
 &\int_{\mathcal C}\dd x\,\gamma_k(x)
       \sum_n\int_{\mathcal C}\dd x'\,F(x',n\mid x)G(x,x',n)O(x')
       \nonumber\\
 &\hspace{2em}=\int_{\mathcal C}\gamma_{k+1}(x')O(x')\dd x'.
 \label{eq:torus-identity}
\end{align}
For the Gaussian update, the kernel ratio is
\begin{equation}
 \log\frac{b(x\mid u)}{f(u\mid x)}
 =\frac{\|u-x-v_ks_\theta(x,t_k)\|^2-\|x-u\|^2}{2v_k}.
 \label{eq:gaussian-ratio}
\end{equation}
Equation~\eqref{eq:gaussian-ratio} uses the unwrapped coordinate actually generated.
If the integer lift is marginalized out, the angular transition kernel must sum over all periodic images.

For $U(1)$ the initial density $q_0$ is the uniform (Haar) measure on the link angles. The initial factor $\gamma_0/q_0$ and the endpoint factor $e^{-S_\beta}/\gamma_T$ connect Eq.~\eqref{eq:torus-identity} to the physical target.
Inserting a kernel invariant under the corresponding normalized $\gamma_k$ leaves the target marginal unchanged.
Resampling uses unbiased offspring counts.
Finite-particle bias and variance of normalized observables are assessed through the independent runs of Sec.~\ref{sec:cost}.

\subsection{Effective coupling of the physics term}
\label{app:weakfield}
The coupling $\beta_{\rm eff}(t)$ of Eq.~\eqref{eq:u1_beta_eff} follows from the variance of one plaquette angle. In the weak field limit the plaquette angles of the target are Gaussian with variance $1/\beta$. A plaquette contains four links, so independent link noise of variance $\sigma^2$ adds variance $4\sigma^2$. The noisy plaquette angle therefore has variance $1/\beta+4\sigma^2$, and the coupling of a Gaussian with this variance is $\beta/(1+4\beta\sigma^2)$. Adjacent plaquettes share links, so their noise is correlated. The physics term ignores these correlations. Section~S2 of the Supplemental Material gives the resulting momentum dependence of the smoothed force.

\subsection{Construction for two-dimensional \texorpdfstring{$SU(3)$}{SU(3)}}
\label{app:su3-construction}
For $SU(3)$ the links are updated in the Lie algebra. With generators $T_a=i\lambda_a/2$, where $\lambda_a$ are the Gell-Mann matrices, an algebra vector $\eta\in\R^8$ defines $X(\eta)=\sum_a\eta_aT_a$. The initial density $q_0$ is the Haar measure. One step of the sampler draws $\eta$ for every link from a Gaussian with mean $g^2(t)\Delta t\,s_\theta$ and variance $g^2(t)\Delta t$ per component and sets $U\to\exp[X(\eta)]\,U$. The noise variance per algebra component is $\tau(t)=\sigma_{\max}^2(1-t)^2$ with $\sigma_{\max}=3$, and $g^2(t)=-\dd\tau/\dd t$.

The intermediate log density at time $t$, with $\gamma_k=\gamma_{t_k}$, is
\begin{equation}
\log\gamma_t(U)=\frac{\beta_{\rm eff}(t)}{3}\sum_z\operatorname{Re}\operatorname{Tr}U_\square(z)+\mathcal E_\theta(U,t;\beta),
\label{eq:su3-bridge}
\end{equation}
\begin{equation}
\beta_{\rm eff}(t)=\frac{\beta\,e^{-8\tau(t)/3}}{1+2\beta\tau(t)/3}.
\label{eq:su3-beta-eff}
\end{equation}
Each factor follows from one limit. At small $\beta$ the heat kernel smooths each link by $e^{-2\tau/3}$ in the fundamental character, and the four links of a plaquette give $e^{-8\tau/3}$. At large $\beta$ each algebra component of the plaquette is Gaussian with variance $6/\beta$, the four links add $4\tau$, and the coupling of the noisy plaquette is $\beta/(1+2\beta\tau/3)$, as in Appendix~\ref{app:weakfield}. The product reduces to the first form for $\beta\tau\ll1$ and to the second for $\tau\ll1$, and we use it as the physics term at all times. The path weights are exact for any such choice. The residual $\mathcal E_\theta$ is a sum over sites of network outputs built from the traces of plaquettes, of $2\times1$ rectangles, and of the clover field strength. These inputs are gauge invariant, and the residual is made even under charge conjugation. The drift $s_\theta$ has the components $\partial_{\eta_a}\log\gamma_t(e^{X(\eta)}U)$ at $\eta=0$.

The path weights use the density of the Gaussian step pushed forward by the exponential map, relative to the Haar measure. This density contains the Jacobian $\prod_{j<k}\operatorname{sinc}^2[(\varphi_j-\varphi_k)/2]$ of the eigenphases $\varphi_j$ of $X$ and a sum over their periodic images, in analogy to the integer lifts of the $U(1)$ angles. The backward kernel is the drift-free Gaussian, and the last increment connects $\gamma$ at the final step directly to the target $e^{-S_\beta}$. After each resampling, $4$ HMC moves with $10$ leapfrog steps act with the potential $-\log\gamma_t$. Their step size is scaled by $\max\{\beta_{\rm eff}(t),1\}^{-1/2}$, adapted to an acceptance of $0.9$ in a pilot run, and then frozen. The model is trained at $L=8$ for $40$ rounds at $\beta\in\{6,8,10,12,14,16\}$. Its training configurations are endpoints corrected by resampling and $8$ HMC moves with the exact action, and the regression target is the exact force of Eq.~\eqref{eq:astra_physical_force}.

\section{Finite-volume references}
\label{app:reference}
The main text compares all results for gauge fields with exact values on the same finite periodic lattice. This appendix gives the formulas.

\subsection{Two-dimensional \texorpdfstring{$U(1)$}{U(1)}}
All exact two-dimensional $U(1)$ values used in the main text follow from the formulas of this subsection.
The reference results follow from the finite-volume solution of periodic two-dimensional pure gauge theory~\cite{Kanwar2020,Albandea2021}.
For a contractible rectangular loop of $A=rs$ plaquettes, with positive integers $r,s$,
\begin{equation}
 \langle W_{r\times s}\rangle
 =\frac{\displaystyle\sum_{n\in\Z}
             I_n(\beta)^{V-A}I_{n+1}(\beta)^A}
        {\displaystyle\sum_{n\in\Z}I_n(\beta)^V}.
 \label{eq:exact-loop}
\end{equation}
Here $I_n$ is the modified Bessel function of the first kind.
The integer $n$ labels the character expansion.
It is distinct from the geometric charge $Q$, which is defined by the total principal-value flux of a configuration.

For the probabilities of geometric sectors, define the density on the real line
\begin{equation}
 f_\beta(u)=
 \frac{e^{\beta\cos u}}{2\pi I_0(\beta)}\,
       \mathbf1_{(-\pi,\pi]}(u).
\end{equation}
Imposing the constraint on the total flux of the periodic lattice gives
\begin{equation}
 P_{\beta,V}(Q=q)
 =\frac{f_\beta^{*V}(2\pi q)}
        {\displaystyle\sum_{m\in\Z}f_\beta^{*V}(2\pi m)}.
 \label{eq:exact-sectors}
\end{equation}
The notation $f_\beta^{*V}$ denotes the $V$-fold convolution on the real line.
The total angle $2\pi q$ separates topological sectors.
Writing $P(q)=P_{\beta,V}(Q=q)$,
\begin{align}
 \chi_{\mathrm{top}}&=\frac1V\sum_q q^2P(q),\\
 \Var(Q^2)&=\sum_q q^4P(q)
                    -\left[\sum_q q^2P(q)\right]^2 .
\end{align}
Charge conjugation gives $\langle Q\rangle=0$.
The variance enters the observable effective sample size in Eq.~\eqref{eq:obs-efficiency}.

\subsection{Two-dimensional \texorpdfstring{$SU(3)$}{SU(3)}}
\label{app:su3-reference}
The exact values for $SU(3)$ follow from the character expansion on the torus. For an irreducible representation $\rho$ with dimension $d_\rho$ and character $\chi_\rho$, let $x_\rho=d_\rho^{-1}\int\dd U\,e^{(\beta/3)\operatorname{Re}\operatorname{Tr}U}\chi_\rho(U)^*$. On a periodic lattice of $V$ plaquettes
\begin{equation}
Z=\sum_\rho x_\rho^V,\qquad
\langle W_A\rangle=\frac1{3Z}\sum_\rho\sum_{\rho'\in\rho\otimes\mathbf 3}\frac{d_{\rho'}}{d_\rho}\,x_\rho^{V-A}x_{\rho'}^A,
\label{eq:su3-exact}
\end{equation}
where $\mathbf 3$ is the fundamental representation and $A$ is the area of a contractible loop. In two dimensions $\langle W_{r\times s}\rangle$ depends only on $A=rs$, and the plaquette is $\langle W_{1\times1}\rangle$. The coefficients $x_\rho$ are computed by Weyl integration over the eigenphases, with Dynkin labels up to $24$. Two independent implementations agree, and a heat bath simulation at $L=4$ reproduces them.

\section{Sampling algorithm and statistics}
\label{app:algorithms}
\subsection{Sampling with a frozen model}
\label{app:production}

After training, the model parameters are fixed. Algorithm~\ref{alg:production} in Sec.~\ref{sec:astra_production} lists the operations of a sampling run.

All weights are computed in the log domain with stable normalization.

The calculations in the main text run these algorithms with the following parameters. The calculations at large volume and along lines of constant physics use $N=1024$ or $4096$ as stated in the captions, $T=4096$, $\sigma_{\max}=2$, and an ESS threshold of $1/2$.
Each resampling step is followed by $4$ HMC moves, each with $15$ leapfrog position steps.
The HMC step size is
\begin{equation}
 \epsilon_{\rm HMC}(t)
 =0.12\sqrt{\frac{32}{L}}\,
                  \frac1{\sqrt{\max\{\beta_{\rm eff}(t),1\}}}.
\end{equation}

The calculations along lines of constant physics use a model trained only at $L=8$ for $150$ rounds at couplings $\{6,13.5,24,37.5,54,96\}$.
All $U(1)$ results of the main text use this model. Couplings between the training values are reached through the conditioning on $\beta$.
For these $U(1)$ models, the intermediate density is Eq.~\eqref{eq:learned-bridge}, including the residual $\mathcal E_\theta$ of the log density.
Weights, HMC potentials, and drifts are all computed from this function.

\begin{figure*}[!tp]
\centering\includegraphics[width=\textwidth]{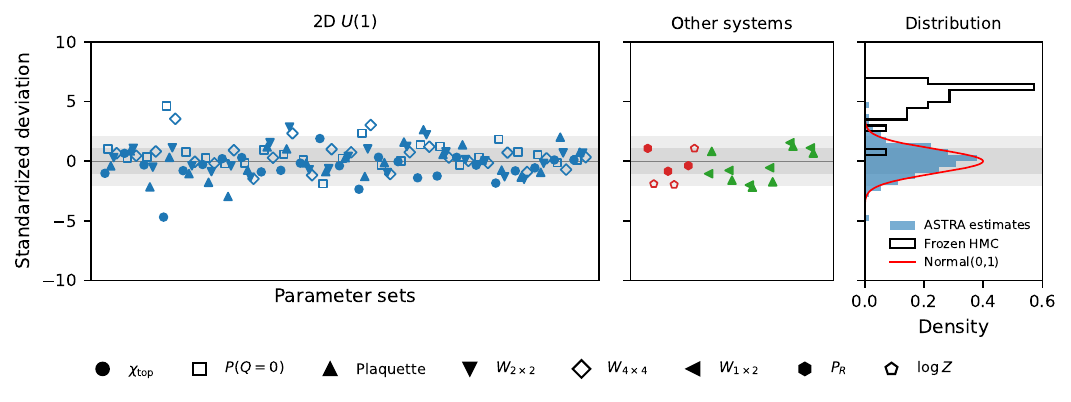}
\caption{Standardized deviations $(\overline O-O_{\mathrm{ref}})/\mathrm{SE}(\overline O)$ from reference observables. The first two panels show individual comparisons for two-dimensional $U(1)$ (125 entries) and other systems (18 entries). The blue markers denote $U(1)$, green two-dimensional $SU(3)$, and red the double well, where the entries include the logarithm of the normalizing constant, $\log Z$. The shaded bands indicate one and two standard errors. The last panel compares the distribution with a standard normal density and the deviations of the topological susceptibility in frozen reference HMC chains with a hot start.}
\label{fig:U1_standardized_deviations}
\end{figure*}

\subsection{Empirical weight variance within and between sectors}
\label{app:resampling}\label{app:sector-variance}

The distribution of weight across sectors affects the ESS. This subsection splits the empirical weight variance into parts between and within sectors.
Let an endpoint pool contain $M$ particles, distinguishing its size from the $N$ particles in a single run.
For positive weights $w_i$, set $u_i=M\bar w_i$, where $\bar w_i=w_i/\sum_lw_l$, so that the mean of $u_i$ is one.
Group particles by $J_i=|Q(x_i)|$.
Let $M_j$ be the count in group $j$ and $\bar u_j=M_j^{-1}\sum_{i:J_i=j}u_i$.
The contributions between and within groups are
\begin{align}
 D_{\rm between}&=\sum_j \frac{M_j}{M}(\bar u_j-1)^2,\nonumber\\
 D_{\rm within}&=\frac1M\sum_j\sum_{i:J_i=j}(u_i-\bar u_j)^2.
 \label{eq:empirical-sector-parts}
\end{align}
The decomposition of the sum of squares gives
\begin{equation}
 D_{\rm between}+D_{\rm within}
 =\frac1M\sum_i(u_i-1)^2
 =\frac{M}{\ESS_w}-1.
 \label{eq:empirical-sector-variance}
\end{equation}
The sum runs over the $|Q|$ groups occupied in the pool.
At $L=16$ and $\beta=24$, averaged over eight such pools of full ASTRA, the part between sectors is about $0.0006$ and the part within sectors is about $0.41$.

\subsection{Ancestors within a sector}
\label{app:ancestors}

The ESS describes only the final weights. Ancestor labels describe where the particles of a sector came from.
Let $\mathrm{anc}(i)$ be the label of the initial ancestor of particle $i$ and $Q_i=Q(x_T^{(i)})$ its endpoint charge.
The number of distinct ancestors in signed sector $Q=j$ is
$N_{\mathrm{anc}}(j)=\#\{\mathrm{anc}(i):Q_i=j\}$.
One ancestor can have descendants in several sectors, so $\sum_jN_{\mathrm{anc}}(j)$ generally differs from the total number of distinct ancestors.
The number of initial ancestors with surviving descendants is $N_{\rm anc}=\#\{\mathrm{anc}(i)\}$ over all final particles. Section~S4 of the Supplemental Material gives its dependence on the volume.
The $|Q|$ distribution among descendants measures topological variation within a family.
Charges before and after accepted HMC moves measure the direct HMC contribution.

\subsection{Statistics of independent runs}
\label{app:run-stats}

The quantities above describe the particle population of one run. Statistical errors of observables come from differences between independent runs, and this subsection defines the statistics used.
For $R$ independent runs with estimates $\widehat O_j$, define
\begin{align}
 \overline O&=\frac1R\sum_{j=1}^{R}\widehat O_j,\nonumber\\
 s_O^2&=\frac1{R-1}\sum_{j=1}^{R}(\widehat O_j-\overline O)^2,\nonumber\\
 \operatorname{SE}(\overline O)&=s_O/\sqrt R.
 \label{eq:run-stats}
\end{align}
Particles within a population can share ancestors.
The independent run is therefore the independent statistical unit.
Table~\ref{tab:imh} shows why. An IMH chain over the final particles of one run gives $N_{\rm eff}^{\rm IMH}/N\approx0.4$ at every $L$. The variation between runs gives $N_{\rm eff}[\chi_{\rm top}]/N$ that falls from $0.45$ at $L=8$ to $0.053$ at $L=32$, together with $N_{\rm anc}$. A diagnostic within one population therefore misses the correlations created by resampling.

\begin{table}[!htbp]
\centering
\caption{IMH chains over the final particles compared with independent runs on the line $\beta/L^2=3/32$ ($N=1024$). $\tau_{\rm int}$ is in IMH steps (Gamma method of pyerrors with window parameter $S=2$), averaged over $8$ runs per point. The IMH estimate $N_{\rm eff}^{\rm IMH}/N$ is $1/[2\tau_{\rm int}(Q^2)]$. The last two columns are from $18$ independent runs.}
\label{tab:imh}
\setlength{\tabcolsep}{3pt}
\resizebox{\columnwidth}{!}{%
\begin{tabular}{rcccccc}
\toprule
$L$ & $\tau_{\rm int}(Q^2)$ & $\tau_{\rm int}(W_{1\times1})$ & $\tau_{\rm int}(W_{4\times4})$ & $N_{\rm eff}^{\rm IMH}/N$ & $N_{\rm eff}[\chi_{\rm top}]/N$ & Ancestors\\
\midrule
8 & 1.02 & 1.14 & 1.14 & 0.49 & 0.45 & 699\\
12 & 1.07 & 1.07 & 1.11 & 0.47 & 0.40 & 351\\
16 & 1.36 & 1.35 & 1.41 & 0.37 & 0.50 & 206\\
20 & 1.08 & 1.05 & 1.01 & 0.46 & 0.17 & 131\\
24 & 1.39 & 1.21 & 1.28 & 0.36 & 0.14 & 89\\
32 & 1.12 & 1.10 & 1.10 & 0.45 & 0.053 & 45\\
\bottomrule
\end{tabular}}
\end{table}
\subsection{Accuracy across all comparisons}
\label{app:pulls}

The standardized deviation $z_O$ is defined in Sec.~\ref{sec:u1_accuracy}. Figure~\ref{fig:U1_standardized_deviations} summarizes these deviations for the topological susceptibility, sector probabilities, plaquette, and Wilson loops, together with results from other models, including $\log Z$ for the double well. Among the 143 comparisons, 61\% satisfy $|z_O|<1$ and 90\% satisfy $|z_O|<2$.
\subsection{Replicates}
\label{app:replicates}
The double well uses five training seeds per setting in Fig.~\ref{fig:double_well} and ten in Table~\ref{tab:dw_weights}, with $16$ sampling runs per frozen model. The error bars of Fig.~\ref{fig:double_well} are standard errors over the five models. On the two $U(1)$ lines of constant physics each point has $18$ runs with $N=1024$ and $20$ runs with $N=4096$. Sampling ablations use $16$ runs per setting, and two-dimensional $SU(3)$ uses $64$ runs per point. Every run has its own random seed. The data of each run, with its checkpoint, seed, particle number, and path length, are in the repository given in the Data and code availability section.

\bibliography{prest}

\clearpage
\setcounter{section}{0}\setcounter{subsection}{0}\setcounter{equation}{0}
\setcounter{figure}{0}\setcounter{table}{0}\setcounter{algorithm}{0}
\renewcommand{\thesection}{S\arabic{section}}\renewcommand{\theHsection}{S\arabic{section}}
\renewcommand{\theequation}{S\arabic{equation}}\renewcommand{\theHequation}{S\arabic{equation}}
\renewcommand{\thefigure}{S\arabic{figure}}\renewcommand{\theHfigure}{S\arabic{figure}}
\renewcommand{\thetable}{S\arabic{table}}\renewcommand{\theHtable}{S\arabic{table}}
\renewcommand{\thealgorithm}{S\arabic{algorithm}}\renewcommand{\theHalgorithm}{S\arabic{algorithm}}
\makeatletter
\def\@hangfrom@section#1#2#3{\@hangfrom{#1#2}\MakeTextUppercase{#3}}%
\let\@sectioncntformat\relax
\@removefromreset{equation}{section}
\def\thesubsection{\Alph{subsection}}\def\thesubsubsection{\arabic{subsubsection}}
\makeatother
\onecolumngrid
\begin{center}{\large\bfseries Supplemental Material}\end{center}
\vspace{1em}
\twocolumngrid
{This Supplemental Material gives derivations and results that support the main text.}

\section{Regression target and training distribution}
\label{app:regression-target}

We write $\sigma=\sigma(t)$ for the noise scale of the main text. A training configuration $X$ is drawn from the buffer $\mathcal B$, and the noise kernel $\mathcal K_\sigma(Y\mid X)$ generates $Y$. The regression label is the physical force $s_\beta(X)=-\nabla S_\beta(X)$. If $X$ has distribution $r$, the squared loss is minimized by
\begin{equation}
 m_{r,\sigma}(y)=\E_r[s_\beta(X)\mid Y=y].
 \label{eq:population-regression}
\end{equation}
When $r=p_\beta$, this is the score of the noisy target density $p_{\beta,\sigma}$,
\begin{equation}
 m_{p_\beta,\sigma}(y)=\nabla_y\log p_{\beta,\sigma}(y).
 \label{eq:tsi}
\end{equation}
This target score identity follows from integration by parts~\cite{DeBortoli2024}. For periodic densities and noise kernels the boundary terms cancel, so it also holds for angular variables. The model fits this conditional mean within its family of gradient fields.

\subsection{Loss for a finite buffer}
\label{app:regression}
Let the buffer contain configurations $x_b$ with sampling probabilities $a_b\ge0$ and $\sum_ba_b=1$. The loss normalized by the lattice volume $V$ is
\begin{equation}
 \mathcal L_{\mathcal B}(s)=\frac1V\sum_b a_b\int \mathcal K_\sigma(y\mid x_b)
                 \|s(y)-s_\beta(x_b)\|^2\dd\mu(y).
 \label{eq:buffer-risk}
\end{equation}
Bayes' rule gives its regression target,
\begin{equation}
 m_{\mathcal B,\sigma}(y)=
 \frac{\sum_b a_b\mathcal K_\sigma(y\mid x_b)s_\beta(x_b)}
      {\sum_b a_b\mathcal K_\sigma(y\mid x_b)}.
 \label{eq:empirical-regression}
\end{equation}

\subsection{Shift from a mismatched buffer}
\label{sm:bound}
Let $\rho=r/p_\beta$. Given $Y=y$, the conditional distribution of $X$ under the target is
\begin{equation}
 \nu_y(\dd x)=\frac{p_\beta(x)\mathcal K_\sigma(y\mid x)}{p_{\beta,\sigma}(y)}\dd\mu(x).
 \label{eq:conditional-target}
\end{equation}
Where $p_{\beta,\sigma}(y)>0$ and $0<\E_{\nu_y}\rho<\infty$, reweighting gives $m_{r,\sigma}=\E_{\nu_y}[\rho\,s_\beta]/\E_{\nu_y}\rho$. Subtracting $m_{p_\beta,\sigma}=\E_{\nu_y}s_\beta$ gives
\begin{equation}
 m_{r,\sigma}(y)-m_{p_\beta,\sigma}(y)
 =\frac{\Cov_{\nu_y}(s_\beta,\rho)}{\E_{\nu_y}\rho}.
 \label{eq:buffer-shift}
\end{equation}
The shift is the conditional covariance of the force and the density ratio. Suppose the buffer mainly changes the relative probabilities of two wells and approximates the distribution inside each well. At noise levels that still identify the well, $\rho$ is nearly constant under $\nu_y$, and the shift is small. Where the noisy wells overlap, different density ratios and forces enter the same conditional mean, and the shift is nonzero. The noise scale therefore controls when information on mode probabilities enters force regression.

\section{Weak field response and Wilson loops at fixed charge}
\label{sm:weakfield}

\begin{figure*}[!tp]
\centering\includegraphics[width=\textwidth]{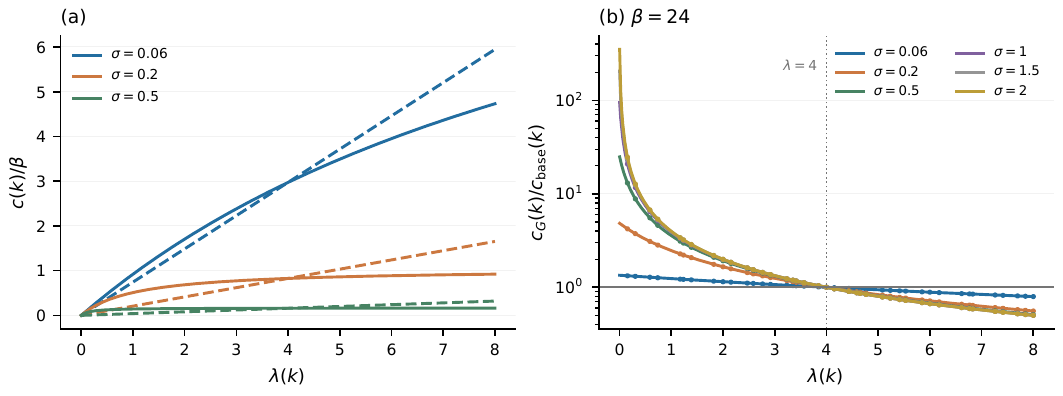}
\caption{Momentum dependence of the weak field recovery coefficients in Eq.~\eqref{eq:response}. (a) Gaussian $c_G/\beta$ (solid lines) and physics term $c_{\mathrm{base}}/\beta$ (dashed lines), with $\beta=24$ and $\sigma=0.06$, $0.2$, $0.5$. (b) Their ratio at $\beta=24$ for $\sigma$ from $0.06$ to $2$, equal to $1$ at $\lambda=4$ (dotted line). Dots mark the momenta of an $L=16$ lattice.}
\label{fig:weakfield}
\end{figure*}

Near $Q=0$ the weak field limit can be solved in closed form. We compute the smoothed score of the Gaussian field and its recovery coefficient $c(k)$ at each momentum $k$ (Fig.~\ref{fig:weakfield}). The weak field action is
\begin{equation}
 S_G(x)=\frac\beta2 x^{\mathsf T}C^{\mathsf T}Cx.
 \label{eq:gaussian-action}
\end{equation}
On $\mathcal H=\operatorname{range}(C^{\mathsf T})$, the operator $C^{\mathsf T}C$ is positive definite.
The covariance is $\Sigma=[\beta C^{\mathsf T}C]_{\mathcal H}^{-1}$.
Add noise with covariance $\sigma^2 I_{\mathcal H}$, where $I_{\mathcal H}$ is the identity on $\mathcal H$.
Then
\begin{align}
 s_G^\sigma(y)
 &=-(\Sigma+\sigma^2I_{\mathcal H})^{-1}y\nonumber\\
 &=-\beta C^{\mathsf T}
       (I+\beta\sigma^2CC^{\mathsf T})^{-1}Cy .
\end{align}
The final equality follows mode by mode for nonzero singular values.
On a periodic square lattice, $CC^{\mathsf T}$ is the plaquette Laplacian.
For nonzero momentum $k_\mu=2\pi n_\mu/L$, with integer $n_\mu$, define $s(k)=-c(k)x(k)$.
The responses are
\begin{align}
 \lambda(k)&=4\sum_{\mu=0}^{1}\sin^2(k_\mu/2),\label{eq:lambda}\\
 c_G(k)&=\frac{\beta\lambda(k)}{1+\beta\sigma^2\lambda(k)},\nonumber\\
 c_{\mathrm{base}}(k)&=\frac{\beta\lambda(k)}{1+4\beta\sigma^2}.
 \label{eq:response}
\end{align}

Appendix~\ref{app:weakfield} of the main text derives the physics term from the variance of one plaquette, which ignores the correlations between adjacent plaquettes. The response difference is
\begin{equation}
 c_G-c_{\rm base}
 =\frac{\beta^2\sigma^2\lambda(4-\lambda)}
       {(1+\beta\sigma^2\lambda)(1+4\beta\sigma^2)}.
 \label{eq:response-difference}
\end{equation}
For nonzero momentum and $\sigma>0$, it is positive at $\lambda<4$ and negative at $\lambda>4$.
It vanishes at $\sigma=0$ or $\lambda=4$.

\subsection{Wilson loops at fixed charge}

The previous subsection treats fluctuations inside one topological sector. Here we add the topological charge and compute Wilson loops at fixed $Q$.
Write the principal plaquette angles as
$\bar\vartheta_z=\operatorname{Arg}e^{i\vartheta_z}=2\pi Q/V+\delta\vartheta_z$, with $\sum_z\delta\vartheta_z=0$.
The quadratic action gives
\begin{equation}
 \Cov(\delta\vartheta_z,\delta\vartheta_{z'})
 =\beta^{-1}\left(\delta_{zz'}-\frac1V\right),
\end{equation}
where $\delta_{zz'}$ is the Kronecker delta.
For a contractible region of $A$ plaquettes, the total angle $\Theta_A$ satisfies
\begin{equation}
 \E[\Theta_A\mid Q]=\frac{2\pi QA}{V},\qquad
 \Var(\Theta_A\mid Q)=\frac{A}{\beta}\left(1-\frac AV\right).
\end{equation}
The Gaussian characteristic function then gives
\begin{equation}
 \E[W_A\mid Q]\simeq
 e^{-A(1-A/V)/(2\beta)}\cos(2\pi QA/V).
 \label{eq:conditional-loop}
\end{equation}

If sampling within each sector is adequate, the average of a large loop contains the factor $\sum_QP(Q)\cos(2\pi QA/V)$.
It is therefore sensitive to sector probabilities.
Numerical comparisons use the references for finite lattices of Appendix~\ref{app:reference} of the main text.

\begin{figure*}[!tp]
\centering\includegraphics[width=\textwidth]{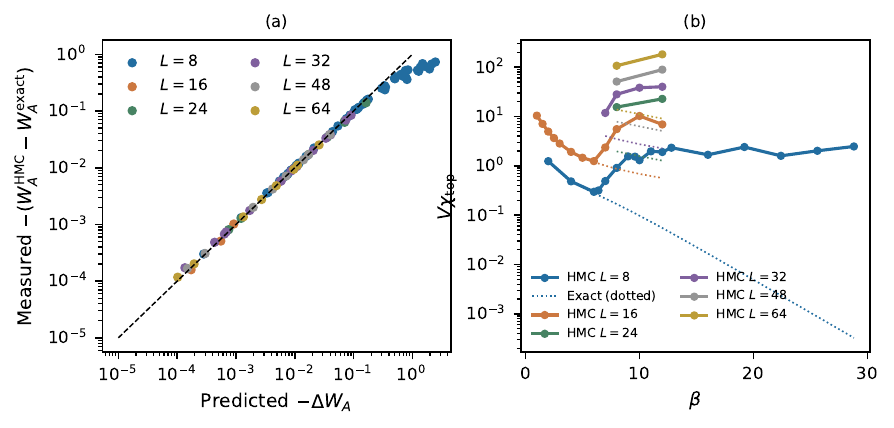}
\caption{Deviations of Wilson loops and topological sampling in reference HMC with a hot start. (a) Measured loop deviations $-(W_A^{\rm HMC}-W_A^{\rm exact})$ of $r\times r$ loops, $r=1$ to $4$, compared with the weak field prediction $-\Delta W_A$, given by Eq.~\eqref{eq:wilson-deficit-prediction} multiplied by $W_A^{\rm exact}$. The dashed line marks equality. The $L=16$ points with predicted deviation below $2\times10^{-5}$ lie at the statistical noise floor and are not shown in (a). (b) $V\chi_{\rm top}=\langle Q^2\rangle$ versus $\beta$. Dotted lines show exact finite-volume values in the color of each $L$. The results use 64 chains with a hot start and 20000 trajectories per chain.}
\label{fig:hmc-wilson-deficit}
\end{figure*}

This sensitivity affects reference Hamiltonian Monte Carlo (HMC), run directly at the target coupling. Cold and hot starts can populate different topological sectors and shift the HMC estimates of Wilson loops.
In the weak field approximation, let $\Delta\langle Q^2\rangle$ be the difference between the reference HMC second moment and its reference.
Equation~\eqref{eq:conditional-loop} gives the leading relative loop deviation,
\begin{equation}
  \frac{\Delta W_A}{W_A^{\rm exact}}
  \simeq-\frac12\left(\frac{2\pi A}{V}\right)^2
  \Delta\langle Q^2\rangle .
  \label{eq:wilson-deficit-prediction}
\end{equation}
Figure~\ref{fig:hmc-wilson-deficit} compares the measured reference HMC deviations with this approximation.
It relates the shifts of large loops in the frozen chains with a hot start to topological sampling.

\section{Training}
\label{sm:training}
Algorithm~\ref{sm:alg:training} lists the operations of one training round, which implement the force regression of Sec.~\ref{sec:astra_training} of the main text.

\begin{algorithm}[!htbp]
\caption{One training round.}
\label{sm:alg:training}
\begin{algorithmic}[1]
\Require current model $\theta$; buffer $\mathcal B_\beta$ at the selected coupling
\State choose a coupling $\beta$ and a generation budget
\State draw independent paths from $q_0$; record the forward and backward Gaussian kernels and the final weights
\State resample endpoints multinomially according to the normalized weights
\State apply the symmetries of the target, then short HMC moves at $\beta$
\State add the configurations to $\mathcal B_\beta$
\For{each training batch $\{X_b\}$ drawn from $\mathcal B_\beta$}
  \State draw a time $t$ and add noise to each $X_b$
  \State update $\theta$ with the label $s_\beta(X_b)=-\nabla S_\beta(X_b)$ and a sampled estimate of Eq.~\eqref{eq:buffer-risk}
\EndFor
\State evaluate generated paths at a fixed budget and save the checkpoint
\end{algorithmic}
\end{algorithm}
The training runs use $4$ to $8$ endpoint HMC moves, Adam, a cosine learning-rate schedule, and gradient clipping at norm $1$.
The repository given in the main text lists the training settings of each checkpoint.

\subsection{Weights in the training loss}
\label{sm:app:reweight}

Endpoint resampling selects configurations according to their weights. The weights can instead enter the loss directly. For a training batch with path weights $w_i$ and an exponent $\eta_k\in[0,1]$ that may grow with the training round $k$,
\begin{align}
 \mathcal L_k(\theta)&=\sum_i\bar w_i^{(\eta_k)}\,\big\|s_\theta(Y_i,t_i)-s_\beta(X_i)\big\|^2,\label{eq:tempered-loss}\\
 \bar w_i^{(\eta)}&=\frac{w_i^{\eta}}{\sum_jw_j^{\eta}}.\nonumber
\end{align}
With $\eta_k=0$ all endpoints count equally, and the training distribution $r$ of Eq.~\eqref{eq:population-regression} is the model distribution. With $\eta_k=1$, $r$ is the target distribution in expectation. A few large weights then dominate the batch while the model is still far from the target. Intermediate values trade this variance against the shift of Eq.~\eqref{eq:buffer-shift}. The ESS-adaptive rule of ATLAS~\cite{Cheng2026}, based on the importance-weight ESS, takes for each batch the largest $\eta$ whose $\ESS_w$ is at least a fraction $\kappa$ of the batch. Table~\ref{tab:dw_weights} of the main text uses $\kappa=0.8$. Our runs use $\eta_k=1$ through resampling, and the following HMC moves separate the duplicated configurations.

\subsection{Endpoint correction in training}

In $U(1)$ at $L=8$, endpoint resampling and HMC moves during training raise the mean $\ESS_w/N$ of independent paths by factors of $1.15$ to $1.46$. This holds at five couplings from $\beta=8$ to $12$, with three trained models per setting (Fig.~\ref{fig:app_u1_training} of the main text).

\begin{figure*}[!tp]
\centering\includegraphics[width=\textwidth]{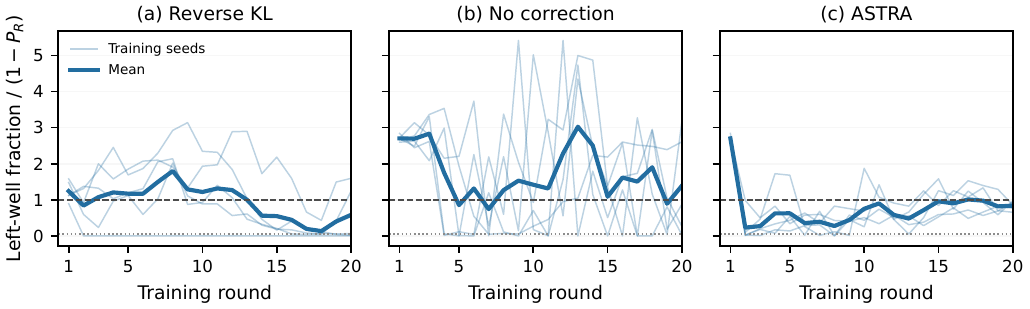}
\caption{Fractions of endpoints in the left well during training, normalized by the exact probability of the left well. (a) Reverse KL in path space. (b) Force regression on unweighted endpoints of independent paths, without resampling or HMC moves (No correction). (c) ASTRA. All panels show generated configurations. Thin lines show the five training seeds, and thick lines show their mean. The dashed horizontal line marks the target fraction, and the dotted line marks the threshold used to identify temporary loss of the left well in a finite sample.}
\label{fig:dw-feedback}
\end{figure*}

In the double well, Fig.~\ref{fig:dw-feedback} follows the fraction of generated configurations, the endpoints of independent paths before weighting and resampling, in the left well during training. It compares reverse Kullback--Leibler (KL) divergence in path space, force regression on unweighted endpoints without resampling or HMC moves, and ASTRA. Three of the five reverse KL models fall below $5\%$ of the exact fraction and never return above $25\%$. All four ASTRA models that fall below $5\%$ return above $25\%$, and so do all five models trained on unweighted endpoints.

\section{Resampling and volume}
\label{sm:resampling}\label{sm:scaling}
Table~\ref{tab:scaling} gives the number of resampling steps per run $n_{\rm rs}$ and the number of initial ancestors with surviving descendants $N_{\rm anc}$ on the two lines of constant physics. All runs use the model trained at $L=8$, with $N=1024$ and $T=4096$. On the first line $n_{\rm rs}/V$ rises from $0.027$ at $L=12$ to $0.036$ at $L=32$, so $n_{\rm rs}$ grows slightly faster than the volume. The variance of the log weight accumulated over a fixed number of steps grows with the volume, so a fixed path length needs more resampling at larger volume. For targets of product form, $\ESS_w$ stays stable when the number of steps grows linearly with the dimension~\cite{Beskos2014}.

For $L\ge16$ the product $N_{\rm anc}n_{\rm rs}/N$ stays between $1.46$ and $1.66$ on both lines, so $N_{\rm anc}\approx1.6\,N/n_{\rm rs}$. This agrees with the genealogy theory of sequential Monte Carlo. Under multinomial and other low-variance resampling schemes and suitable conditions, the genealogies converge to a time-rescaled Kingman coalescent as $N$ grows. Lines of descent then merge at a rate of order $1/N$ per resampling step~\cite{Koskela2020,Brown2021}. Together the two observations give $N_{\rm anc}$ approximately proportional to $N/V$ at fixed physics. On the first line $N_{\rm eff}[\chi_{\rm top}]$ of Table~\ref{tab:imh} of the main text is of the same order as $N_{\rm anc}$ and falls in the same way. 

\begin{table}[!htbp]
\centering
\caption{Resampling steps per run $n_{\rm rs}$ and surviving initial ancestors $N_{\rm anc}$ on the lines $\beta/L^2=3/32$ ($18$ runs per point) and $5/32$ ($10$ runs per point with recorded resampling counts), with $N=1024$ and $T=4096$.}
\label{tab:scaling}
\setlength{\tabcolsep}{4pt}
\begin{tabular}{rrrrrr}
\toprule
$L$ & $\beta$ & $n_{\rm rs}$ & $10^2n_{\rm rs}/V$ & $N_{\rm anc}$ & $N_{\rm anc}n_{\rm rs}/N$\\
\midrule
8 & 6 & 1.0 & 1.6 & 699 & 0.68\\
12 & 13.5 & 3.9 & 2.7 & 351 & 1.33\\
16 & 24 & 7.7 & 3.0 & 206 & 1.54\\
20 & 37.5 & 12.6 & 3.2 & 131 & 1.61\\
24 & 54 & 19.2 & 3.3 & 89 & 1.66\\
32 & 96 & 37.0 & 3.6 & 45 & 1.62\\
\midrule
8 & 10 & 1.9 & 3.0 & 536 & 0.99\\
12 & 22.5 & 4.9 & 3.4 & 292 & 1.39\\
16 & 40 & 9.1 & 3.6 & 164 & 1.46\\
20 & 62.5 & 15.2 & 3.8 & 99 & 1.47\\
24 & 90 & 23.0 & 4.0 & 71 & 1.60\\
\bottomrule
\end{tabular}
\end{table}

\subsection{Scans of the sampling parameters}
\label{sm:scans}
With the model fixed, we vary the path length $T$ and the number of HMC moves after each resampling step, which is $4$ by default. At $L=16$ and $\beta=24$ with $N=1024$, $T=512$, $1024$, $2048$, and $8192$ give $N_{\mathrm{eff}}[\chi_{\rm top}]=39$, $127$, $115$, and $350$. With $2$, $8$, and $16$ HMC moves the values are $343$, $281$, and $370$. At $L=32$ and $\beta=96$, the same path lengths give $4$, $9$, $25$, and $38$, and $2$, $8$, and $16$ HMC moves give $27$, $41$, and $73$. Each of these settings has $10$ runs. The default setting at $L=32$ gives $54$ from $18$ runs. Paths shorter than $T=4096$ give lower values. All estimates of $\chi_{\rm top}$ agree with the exact value within $1.9$ standard errors.

At $L=64$ and $\beta=12$, the same model gives $\overline{\chi}_{\rm top}/\chi_{\rm top}^{\rm exact}=0.998(35)$ from $10$ runs with $N=1024$. Each run keeps $36$ to $45$ surviving ancestors. Untrained ASTRA gives $1.00(10)$ at $L=32$ and $1.28(27)$ at $L=64$. At $L=64$ its runs keep $2$ to $6$ ancestors, and its standard error is about eight times larger.

\section{Two-dimensional \texorpdfstring{$\phi^4$}{phi4} theory}
\label{sm:dw-phi4}

\label{app:phi4-plan}
The double well has a single multimodal degree of freedom. This section applies the same method to two-dimensional $\phi^4$ theory, where the two branches of the magnetization play the role of the wells.
\begin{table}[!tbp]
\centering
\caption{Two-dimensional $\phi^4$ theory with models trained at $L=8$, $N=1024$, $T=1024$. The upper block gives $\langle|m|\rangle$ at $h=0$, and the lower blocks give $P_+$ and $\langle|m|\rangle$ at $h=0.002$. Parentheses give standard errors over independent runs. Reference errors below the last digit are omitted. Unlike $z_O$ of the main text, $z$ is the deviation in units of the combined standard error, computed from unrounded values.}
\label{tab:phi4}
\setlength{\tabcolsep}{2.5pt}
\begin{tabular}{rrrlll}
\toprule
\multicolumn{6}{l}{$h=0$, $\langle|m|\rangle$}\\
$L$ & $\kappa$ & Runs & ASTRA & Reference & $z$\\
\midrule
32 & 0.26 & 10 & $0.1250(14)$ & $0.1255(16)$ & $-0.3$\\
32 & 0.27 & 10 & $0.619(13)$ & $0.632(12)$ & $-0.7$\\
32 & 0.28 & 10 & $1.4554(11)$ & $1.4555(6)$ & $-0.1$\\
32 & 0.29 & 10 & $1.8126(2)$ & $1.8121(3)$ & $+1.1$\\
32 & 0.30 & 10 & $2.0879(2)$ & $2.0873(2)$ & $+1.6$\\
\midrule
\multicolumn{6}{l}{$h=0.002$, $P_+$ (upper), $\langle|m|\rangle$ (lower)}\\
$L$ & $\kappa$ & Runs & ASTRA & Reference & $z$\\
\midrule
8 & 0.28 & 16 & $0.657(37)$ & $0.5915$ & $+1.7$\\
8 & 0.30 & 16 & $0.704(48)$ & $0.6305$ & $+1.5$\\
16 & 0.28 & 16 & $0.814(35)$ & $0.8162(2)$ & $-0.1$\\
16 & 0.30 & 16 & $0.837(38)$ & $0.8945$ & $-1.5$\\
32 & 0.28 & 10 & $0.981(7)$ & $0.99743(1)$ & $-2.2$\\
32 & 0.30 & 10 & $0.9998(2)$ & $0.99981$ & $+0.2$\\
\midrule
8 & 0.28 & 16 & $1.4481(25)$ & $1.4478(12)$ & $+0.1$\\
8 & 0.30 & 16 & $2.0878(10)$ & $2.0883(3)$ & $-0.4$\\
16 & 0.28 & 16 & $1.4600(15)$ & $1.4612(13)$ & $-0.6$\\
16 & 0.30 & 16 & $2.0892(8)$ & $2.0901(4)$ & $-1.0$\\
32 & 0.28 & 10 & $1.4630(8)$ & $1.4637(6)$ & $-0.7$\\
32 & 0.30 & 10 & $2.0906(5)$ & $2.0904(2)$ & $+0.3$\\
\bottomrule
\end{tabular}
\end{table}

\begin{figure*}[!tp]
\centering
\includegraphics[width=\textwidth]{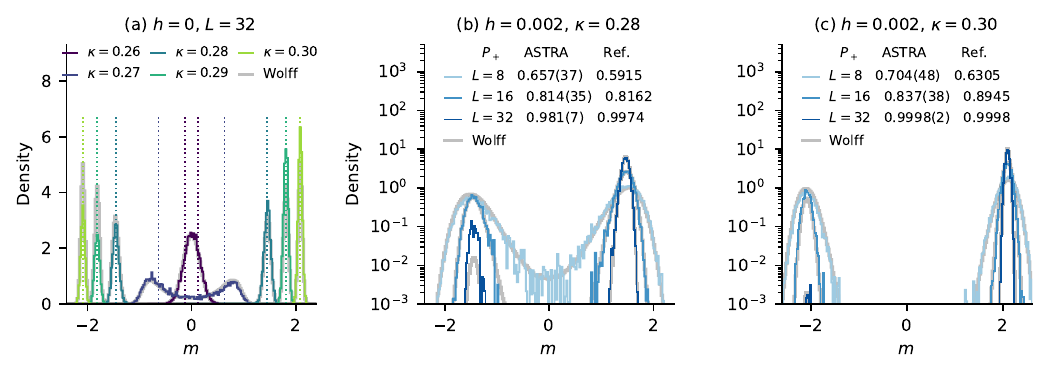}
\caption{Weighted distributions of the magnetization $m$ in two-dimensional $\phi^4$ theory from ASTRA runs with models trained at $L=8$, averaged over the per-run histograms. (a) At $h=0$ and $L=32$ the distribution splits into two branches as $\kappa$ increases, and dotted lines mark $\pm\langle|m|\rangle$ of the reference restricted to $m>0$. (b, c) At $h=0.002$ the same model is applied at $L=8$, $16$ and $32$ for $\kappa=0.28$ and $0.30$. The legends give $P_+$ from ASTRA with the standard error over runs and the reference value, as in Table~\ref{tab:phi4}. Gray curves are independent Wolff cluster simulations, reweighted to $h=0.002$ in (b) and (c).}
\label{fig:phi4}
\end{figure*}

For a periodic square lattice, use the action
\begin{align}
 S[\varphi]&=\sum_x\left[\varphi_x^2+\lambda_4(\varphi_x^2-1)^2\right]\nonumber\\
 &\quad-2\kappa\sum_{x,\mu=1}^{2}\varphi_x\varphi_{x+\hat\mu}
       -h\sum_x\varphi_x .
 \label{eq:phi4}
\end{align}
Here $\lambda_4$ is the quartic coupling, and each nearest-neighbor bond in the positive direction is counted once.
The magnetization is $m=V^{-1}\sum_x\varphi_x$.

For each $\kappa$, a model is trained at $L=8$ with $\lambda_4=0.022$ and applied without retraining at $L=16$ and $32$.
Each run uses $N=1024$ particles and $T=1024$ path steps.
The learned residual is symmetrized under $\varphi\to-\varphi$, and the field term enters exactly.
Table~\ref{tab:phi4} compares $P_+=\Pr(m>0)$ and $\langle|m|\rangle$ with reference values. Figure~\ref{fig:phi4} shows the weighted distributions of $m$.
At $h=0$, $\langle|m|\rangle$ is the mean of $m$ in the $m>0$ branch, and its reference is HMC restricted to $m>0$.
At $h=0.002$, $P_+$ and $\langle|m|\rangle$ follow from the same restricted chains by reweighting with $e^{\pm hVm}$.
All $17$ deviations are below $2.5$ standard errors. Independent simulations with Wolff cluster updates embedded in Metropolis sweeps~\mbox{\cite{Brower1989}} agree with the restricted HMC reference within $0.7$ standard errors at all $15$ common points and give the same deviations of ASTRA.
At $L=32$ the runs retain between one and about twenty initial ancestors (one or two at $\kappa=0.30$), so the errors are taken across independent runs. With so few ancestors, as at $h=0$ and $\kappa\geq0.29$, where runs keep one to five, the spread of ten runs can underestimate the error, and these two estimates lie $1.3$ and $1.7$ standard errors above the Wolff values.

\end{document}